\documentclass[11pt,twoside]{article}

\usepackage[usenames,x11names,table]{xcolor}
\usepackage[font={rm,small,sl}]{caption}
\usepackage{graphicx}
\usepackage{cancel}
\usepackage[obeyspaces]{url}
\usepackage[margin=1.1in]{geometry}
\usepackage{enumerate}
\usepackage{amsmath}
\usepackage{amssymb}
\usepackage{subfigure}
\usepackage{multirow}
\usepackage{listings}
\usepackage{algorithm2e}

\definecolor{darkgray}{rgb}{0.6,0.6,0.6}
\definecolor{darkgreen}{rgb}{0.3,0.5,0.3}
\definecolor{mygreen}{rgb}{0.0,0.4,0.0}
\definecolor{mygray}{rgb}{0.5,0.5,0.5}
\definecolor{mymauve}{rgb}{0.58,0,0.82}

\usepackage{listings}

\lstdefinestyle{MYC}{ 			   
  language=C,                      
  backgroundcolor=\color{white},   
  basicstyle=\scriptsize\ttfamily,        
  breakatwhitespace=false,         
  breaklines=true,                 
  captionpos=b,                    
  commentstyle=\color{mygreen},    
  deletekeywords={...},            
  escapeinside={\%*}{*)},          
  extendedchars=true,              
  keepspaces=true,                 
  keywordstyle=\color{blue},       
  otherkeywords={foreach, func, not},           
  numbers=left,                    
  numbersep=10pt,                   
  numberstyle=\tiny\color{black}, 
  rulecolor=\color{black},         
  showspaces=false,                
  showstringspaces=false,          
  showtabs=false,                  
  stepnumber=1,                    
  stringstyle=\color{mymauve},     
  tabsize=2,	                   
  title=\lstname                   
  pagebreak=true
}

\colorlet{color}{cyan!50}

\newcolumntype{L}[1]{>{\raggedright\let\newline\\\arraybackslash\hspace{0pt}}m{#1}}
\newcolumntype{C}[1]{>{\centering\let\newline\\\arraybackslash\hspace{0pt}}m{#1}}
\newcolumntype{R}[1]{>{\raggedleft\let\newline\\\arraybackslash\hspace{0pt}}m{#1}}

\newcommand{\smalltt}[1]{{\texttt{\small #1}}}

\begin{document}


\title{ChainifyDB: How to Blockchainify any Data Management System}


\author{Felix Martin Schuhknecht\qquad Ankur Sharma\qquad Jens Dittrich\qquad Divya Agrawal\\
\\
Big Data Analytics Group\\
Saarland Informatics Campus\\
\url{https://bigdata.uni-saarland.de}}

\date{\today}

\maketitle

\begin{abstract}

Today's permissioned blockchain systems come in a stand-alone fashion and require the users to integrate yet another full-fledged transaction processing system into their already complex data management landscape.
This seems odd as blockchains and traditional DBMSs share large parts of their processing stack. Thus, rather than replacing the established data systems altogether, we advocate to simply `\textit{chainify}' them with a blockchain layer on top.

Unfortunately, this task is far more challenging than it sounds: As we want to build upon \textit{heterogenous} transaction processing systems, which potentially behave differently, we cannot rely on every organization to execute every transaction deterministically in the same way. Further, as these systems are already filled with data and being used by top-level applications, we also cannot rely on every organization being resilient against tampering with its local data.

Therefore, in this work, we will drop these assumptions and introduce a powerful processing model that avoids them in the first place: The so-called \textit{Whatever-LedgerConsensus~(WLC)} model allows us to create a highly flexible permissioned blockchain layer coined \textit{ChainifyDB} that (a)~is centered around bullet-proof database technology, (b)~makes even stronger guarantees than existing permissioned systems, (c)~provides a sophisticated recovery mechanism, (d)~has an up to $6$x~higher throughput than the permissioned  blockchain system Fabric, and (e)~can easily be integrated into an existing heterogeneous database landscape. 

\end{abstract}

\section{Introduction}
\label{sec:introduction}

The vast majority of modern permissioned blockchain systems (PBS)~\cite{untangling_blockchain, bigchaindb, fabric, multichain, quorum, blockchain_meets_database, blockchaindb}, in which all organizations are known at any time, come as stand-alone end-to-end transaction processing systems. 
As a consequence, an organization that wants to utilize blockchain technology is currently forced to add yet another data management system to its infrastructure. However, since this infrastructure typically already consists of various established systems, which are filled with data and used by top-level applications, this approach is extremely troublesome. Data must be migrated, applications must be rewritten, personnel must be retrained --- in general, the integration and maintenance of a new full-fledged system is associated with high costs and immense
effort.

This raises the question, whether it is actually necessary to reinvent the wheel and design blockchain systems in a stand-alone fashion in the first place. Large parts of the transaction processing stack are conceptually shared with traditional database management systems~\cite{blurring_the_lines}. Why not simply \textit{reuse} these parts and build upon them? Precisely, instead of replacing the established database management systems altogether, we advocate to extend them with the missing and desired blockchain functionality. 

Unfortunately, this task is far more challenging than it sounds at first. This has to do with two fundamental differences between stand-alone PBSs and our design.

First, due to their restrictive design, stand-alone PBSs can make the convenient assumption that:

\begin{enumerate}
\item[1)] \textit{Every organization executes every transaction deterministically in  the same way}.
\end{enumerate}

\noindent As every organization runs the very same storage system and the very same transaction processing engine, it can be assumed comfortably that every transaction is interpreted in exactly the same way and results in the same effect across all organizations. 
In contrast to that, in a heterogeneous setup we don't have the luxury to rely on this assumption. As one organization of the network might build upon DBMS~X while another one builds upon DBMS~Y, the exact same transaction could result in different effects across organizations. Reasons for this are manifold: Systems might implement a different interpretation of the SQL standard or of the used data types. 

The second fundamental difference is that, due to their restrictive design, stand-alone PBSs can assume that:

\begin{enumerate}
	\item[2)] \textit{Every organization is resilient against tampering with its local data}. 
\end{enumerate}

\noindent As stand-alone PBSs employ their own dedicated storage system, they fully control all access that is happening to it. Thus, in practice, it is unlikely that the data is corrupted externally.
However, in our design we cannot make this assumption. As we want to utilize arbitrary database systems containing various forms of data, which are accessed from top-level applications aside from our blockchain layer, the chances of  corruption are significantly higher in our case.

In summary, the processing model employed by the vast majority of stand-alone systems is obviously not powerful enough to handle the challenges of our highly flexible design. To understand the precise problem and to come up with a new and more powerful model, let us inspect the general workflow that is currently applied.

\subsection{Order-Consensus-Execute}

The model, that is currently implemented by the vast majority of stand-alone PBSs~\cite{untangling_blockchain, bigchaindb, fabric, multichain, quorum, blockchain_meets_database, blockchaindb} is called \textit{order-consensus-execute (OCE)}.
First, in the order-phase, an order on a batch of transactions is proposed. 
Then, in the consensus-phase, the organizations try to globally agree on this ordered batch using some sort of consensus mechanism. If a consensus is reached, in the final execute-phase, the agreed-upon ordered batch is executed locally by every organization. As a result, all honest organizations reach the same state.

\begin{figure}[htb!]
	\begin{center}
	\includegraphics[page=1, width=.7\linewidth, trim={2cm 18cm 7cm 3cm}, clip]{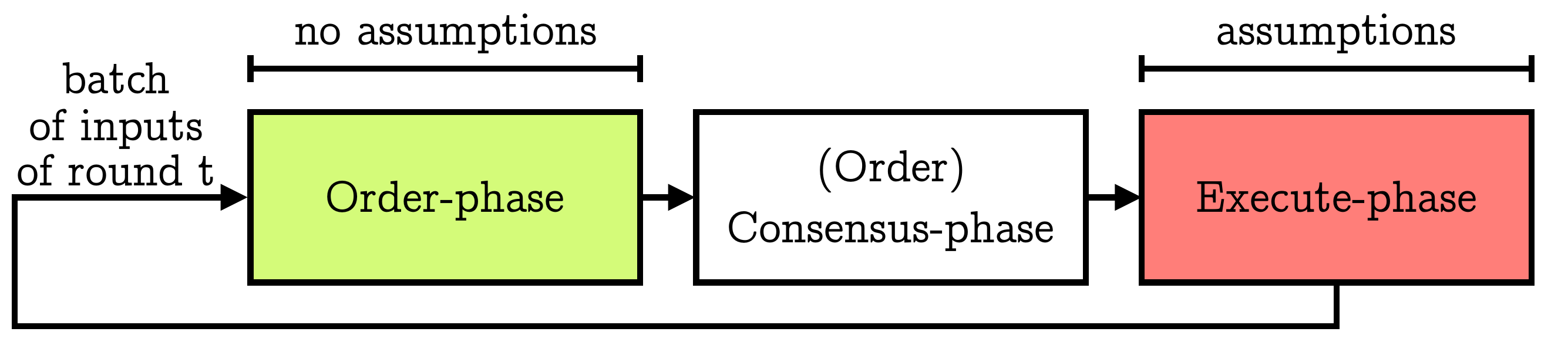}
	\end{center}
	\caption{The order-consensus-execute model (OCE). The consensus-phase sits between the order-phase and the execute-phase. As a consequence of this design, assumptions must be made on everything \textit{after} the consensus-phase, namely on the execute-phase.}
	\label{fig:wlc_concept_olce}
\end{figure}

The problem lies in the assumption, that if a consensus is reached on the result of the order-phase, every honest organization will be in the same state after the execution-phase. In other words, it assumes that everything goes well after the consensus-phase.
Figure~\ref{fig:wlc_concept_olce} visualizes the problem: OCE assumes deterministic behavior on anything happening \textit{after} the consensus-phase, namely on the execute-phase.

Unfortunately, this assumption is not compatible with our design as we want to chainify heterogeneous infrastructures, that potentially behave differently and that are potentially prone to tampering with the data.

\subsection{Whatever-LedgerConsensus}

\begin{figure}[htb!]
	\begin{center}
	\includegraphics[page=1, width=.7\linewidth, trim={2cm 10.5cm 7cm 10.5cm}, clip]{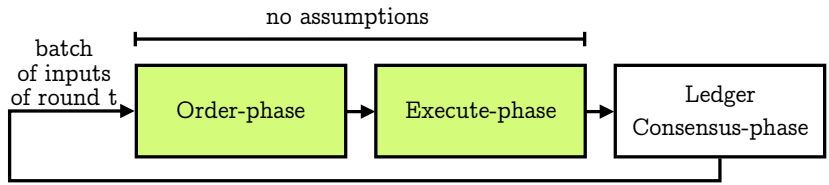}
	\end{center}
	\caption{The order-execute-consensus model (OEC). The consensus-phase sits at the end of the pipeline, after both order-phase and the execute-phase. As consensus is reached on the \textit{effects} of the execute-phase, no assumptions must be made on any previous phase.}
	\label{fig:wlc_concept_oelc}
\end{figure}

To eliminate the need for assumptions, we argue that the consensus-phase must be pushed towards the end of the processing pipeline. Figure~\ref{fig:wlc_concept_oelc} shows the effect of doing so, resulting in the \textit{order-execute-consensus model (OEC)}. If we reach consensus on the effects of the execute-phase instead of reaching consensus on the order established by the order-phase, no assumptions must be made on the order-phase and execute-phase.

From this perspective, we are actually able to abstract from the concrete order-phase and the execute-phase: Whatever happens in these phases, we are able to detect all differences in the produced effects in the consensus-phase afterwards. 

\begin{figure}[htb!]
	\begin{center}
	\includegraphics[page=1, width=.7\linewidth, trim={2cm 3cm 7cm 18cm}, clip]{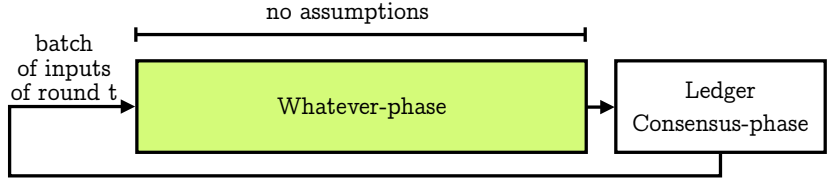}
	\end{center}
	\caption{Our Whatever-LedgerConsensus model (WLC). We do not make assumptions on the behavior of the  whatever-phase. In the consensus-phase, consensus is reached on the effects of the whatever-phase. }
	\label{fig:wlc_concept_wlc}
\end{figure}

This results in a new, highly flexible processing model we call the \textit{Whatever-LedgerConsensus model} or \textit{WLC} for short (pronounced ``We'll see!''). Figure~\ref{fig:wlc_concept_wlc} visualizes the two phases of our WLC model: 

\begin{enumerate}
	\item \textbf{W}hatever-phase. Each organization does whatever it deems necessary to pass the LC-phase later on. 
	\item \textbf{L}edger\textbf{C}onsensus-phase. We perform a consensus round on the effects of the whatever-phase to check whether a consensus can be reached on them. If yes, the effects are committed to a ledger. 
	If an organization is non-consenting, it must perform a recovery. If no consensus is reached at all, all organizations must try to recover.
\end{enumerate}

This WLC model finally allows us to design and implement our highly flexible permissioned blockchain system: \textit{ChainifyDB}. It supports \textit{different} transaction processing systems across organizations to build a \textit{heterogeneous} blockchain network.

\subsection{Contributions}

\begin{enumerate}
\item We present a new processing model coined the \textbf{Whatever-LedgerConsensus model (WLC)}.
In contrast to existing processing models, our model does not make any assumptions on the behavior of the local engines. Still, we are able to detect any deviation of an organization, irrespective of the cause. Our model allows us to realize highly flexible blockchain systems while being able to express existing models like OCE as special cases.

\item We present the \textbf{Whatever Recovery Landscape} and discuss the different classes of recovery algorithms possible in the WLC model, depending on the amount of information available regarding actions and effects. 

\item We present a concrete system instance of WLC coined \textbf{ChainifyDB}. We start with a set of heterogeneous database systems and show how to connect them to a network providing permissioned blockchain-properties. We only requires the DBMSs providing a trigger mechanism as defined in SQL 99 or similar.

\item We show initial results with a \textbf{vendor-independent recovery algorithm} allowing ChainifyDB to efficiently recover non-consenting organizations. Notice that systems like Fabric~\cite{fabric} do not have \textit{any} recovery mechanism.

\item We perform an extensive \textbf{experimental evaluation of ChainifyDB} in comparison with the comparable state-of-the-art permissioned blockchain systems Fabric~\cite{fabric} and Fabric++~\cite{blurring_the_lines} and achieve an up to $6x$~higher throughput. Further, we show that ChainifyDB is able to fully utilize the performance of the underlying database systems and demonstrate its robustness and recovery capabilities experimentally.
\end{enumerate}

The paper is structured as follows. In Section~\ref{sec:related_work}, we first discuss related works, which sit at the intersection of databases and blockchains and contrast WLC and ChainifyDB to them from a conceptual perspective. In Section~\ref{sec:WLC}, we formalize our Whatever-LedgerConsensus model (WLC). In Section~\ref{sec:recovery}, we present the Whatever Recovery Landscape. In Section~\ref{sec:chainifydb}, we present the logical design of ChainifyDB and its components, including the recovery mechanism. In Section~\ref{sec:optimizations}, we present interesting optimizations possible in ChainifyDB. In Section~\ref{sec:arch}, we present concrete implementation details of ChainifyDB. Finally, in Section~\ref{sec:eval}, we perform an extensive experimental evaluation.

\section{Related Work}
\label{sec:related_work}

As stated, the vast majority of permissioned blockchain systems implement variants of the OCE model, with~\cite{ bigchaindb,  multichain, quorum, blockchain_meets_database, blockchaindb} being prominent examples. Even Fabric~\cite{fabric, fabric2}, which uses a model the authors call \textit{execute-order-validate}, implements in the end a form of OCE: In the execute-phase, the effect of transactions are computed. After ordering the transactions, the effects of non-conflicting transactions are used to update the state in the given order. Thus, the validate-phase in Fabric highly resembles the execute-phase in OCE. 
Interestingly, none of these system integrate any form of recovery mechanism: As soon as an organization deviates, it is considered as being malicious independent of the cause and implicitly banned from further transaction processing.

Apart from the execution model, there are other projects that sit at the intersection of databases and blockchains. 
In~\cite{blockchain_meets_database}, the authors extend the relational system PostgreSQL with blockchain features. This results in a ``blockchain relational database'', which is capable of performing trusted transactions between multiple PostgreSQL instances. While this project is clearly a step in the right direction, it does not go far enough: the blockchain relational database still comes in a stand-alone fashion and forces the users to integrate a whole new system into their infrastructure. Further, they heavily modify the internals of PostgreSQL to integrate the blockchain features. In contrast to that, in ChainifyDB we install the blockchain features on-top of blackbox DBMSs without changing them in any way. We intended to compare ChainifyDB against this system, however, the source code is not available.

A project with a similar attitude is BlockchainDB~\cite{blockchaindb, blockchaindb_demo}. In contrast to~\cite{blockchain_meets_database} and to ChainifyDB, the authors of~\cite{blockchaindb, blockchaindb_demo} install a database layer \textit{on top of} an existing blockchain, such as Ethereum~\cite{ethereum} or Hyperledger Fabric~\cite{fabric}. This database layer allows the user to manage and access the underlying shared data in a more convenient way than directly communicating with the blockchain system. Unfortunately, the comfort is relatively limited: To support a variety of blockchain systems, the authors have to stick to a key-value data model and a simple put()/get() query interface. 

Another project at the intersection of databases and blockchains is Veritas~\cite{veritas}. This visionary paper also proposes to extend existing database systems by blockchain features; however, they focus on a cloud infrastructure. To synchronize instances, they utilize log shipping. Therefore, this solution requires the underlying database system to provide log shipping in the first place and disallows the connection of \textit{different} database systems in one network, if their log shipping mechanisms are not compatible with each other. Of course, they are not in general.

The project ChainSQL~\cite{chainsql} takes the open-source blockchain system Ripple~\cite{ripple} and integrates relational and NoSQL databases into the storage backend. This enables it to run SQL-style respectively JSON-style transactions. However, by integrating database systems into the heavy-weight blockchain system Ripple, the authors limit the transaction processing performance to that of Ripple --- thus overshadowing the high performance of the underlying database systems. In contrast to that, ChainifyDB is designed from the get-go to leverage the power of the underlying systems --- in particular using highly parallel transaction execution, even across heterogeneous database systems.

The project BigchainDB~\cite{bigchaindb} combines the blockchain framework Tendermint~\cite{tendermint} with the document store MongoDB and therefore extends it with blockchain features. In contrast to ChainifyDB, the system is shipped in a stand-alone fashion and focuses on the query interface of MongoDB.

Apart from works aiming at closing the gap between databases and blockchains from an architectural perspective, there is a considerable amount of research improving the performance of permissioned blockchains. In~\cite{blurring_the_lines}, \cite{fast_fabric}, \cite{scaling_blockchain}, and~\cite{increasing_concurrency}, the authors apply various optimizations to improve the throughput of successful transactions in Fabric.

\section{Whatever-Ledger Consensus}
\label{sec:WLC}

Let us now start by introducing our novel process model \textit{Whatever-Ledger Consensus (WLC)} and formalize it. 

\subsection{Core Idea}

In short, the core idea of WLC is to \textbf{not} seek consensus on \textit{what should be done}-actions, but rather seek consensus on the effect of the actions \textit{after they have been performed}. 
This allows us to drop assumptions on the concrete transaction processing behavior of the organizations. It also allows us to detect any external tampering with the data. The WLC model implies that if consensus on the effects of certain actions cannot be reached, the organizations must not commit the effects of their actions. 

Note that WLC is also more powerful than classical 2PC or 3PC-style protocols in the sense that they still assume deterministic behavior of the organizations without looking back at the produced effects. 
In WLC, we simply do not care about whether organizations claim to be \textit{prepared} for a transaction and what they claim to \textit{commit}. In contrast, we solely look at the outcome. If we can reach consensus on the outcome, it  does not matter anymore which actions those organizations used to get to the same outcome in the first place. 

In summary, \textit{we measure the outcome rather than the promises}.

\subsection{Processing Model}
\label{ssec:processing_model}

In the following, let us formalize WLC in detail.  

Let $O_1,\ldots,O_k$ be the $k$ participating organizations in the network. As always, these organizations do not trust each other. Still, they want to perform a mutual sequence of inputs, which is fed into the system batch-wise round by round, as visualized in Figure~\ref{fig:wlc_concept_wlc}. The following description shows the W-phase and the LC-phase in round~$t$.

\noindent \textbf{Whatever-phase}.
Note that although the \textit{same} sequence of inputs enters the system, each organization might actually receive a potentially different set of \textit{actions} for processing, as we do not make any assumptions on what exactly is happening in the W-phase. For instance, an ordering service could distribute different actions to different organizations for whatever reason. Thus, we define $A_{l,1},\ldots,A_{l,t}$ as the sequence of actions, that the individual organization~$O_l$ receives till round~$t$.

\noindent\textit{Effect Functions:} Further, we assume that there is an \textit{effect function} $F_l()$. 
Only if the effect contains the state, an action can be applied on the effect to generate a new effect.
We compute for each organization~$O_l$ the accumulated effect~$E_{l,t}$ as:
$$ E_{l,t} = F_l(E_{l,t-1}, [A_{l,t}]) $$
with $E_{l,0}=\emptyset$ being the initial empty effect.
Notice the iterative construction of this function (which will later on create our blockchain-style chaining of effects):
$$ E_{l,t} = F_l(F_l(F_l(\ldots), [A_{l,t-1}]) , [A_{l,t}]). $$
Notice that we use
$E_{l,t} = F_l(\emptyset, [A_{l,1},\ldots,A_{l,t}])\text{,}$ as a shorthand, still there will be a separate effect function call for each of the $t$ actions.

\noindent \textbf{LedgerConsensus-phase}.  
On the accumulated effects~$E_{l,t}$ for $1 \leq l \leq k$, a consensus must be reached (and not necessarily on the entire state!). Otherwise, the system is not allowed to proceed to round~$t+1$.
To decide whether consensus is reached, a \textit{consensus policy}~$c$ specifies how many organization must at least have reached the same accumulated effect.  
Thus, consensus in round~$t$ is reached if 
$$ \exists O_{cons} = \{O_i, ..., O_j\}: |O_{cons}| \ge c: (E_{i,t} = \ldots = E_{j,t}) = E_{cons,t}\text{.}$$
We summarize the effect on which consensus has been reached as $E_{cons,t}$.
If consensus has been reached, each organization~$O_l$ has to decide on its own whether its local effect~ 
$E_{l,t}$ matches the consensus effect~$E_{cons,t}$. If it matches, the effect can be committed to a \textit{ledger} and round~$t$ ends. Otherwise, $O_l$ can not proceed to round~$t+1$ and tries to recover.
If no consensus can be reached at all, i.e. $E_{cons,t}=\text{undefined}$, then no organization can proceed to round~$t+1$. In this case, all organizations try to recover. We will discuss in Section~\ref{sec:recovery} the detailed recovery behavior and introduce a concrete database vendor-independent recovery algorithm in Section~\ref{sec:chainifydb}.

\section{Whatever Recovery}
\label{sec:recovery}

\begin{table*}[ht!]
\footnotesize
\begin{center}
\begin{tabular}{|p{1cm}l||p{2.5cm}|p{3.5cm}|p{3.7cm}|}
\hline
 && \multicolumn{3}{c|}{ \textbf{Actions}} \\
 && \textbf{not accessible} & \textbf{accessible (blackbox)} & \textbf{accessible (whitebox)} \\\hline\hline
\multirow{2}{*}{\textbf{Effects}} & \textbf{State not contained} & --- & + full replay  & + optimized full replay  \\\cline{2-5}
& \textbf{State contained} & + restore state &  + partial replay  & + optimized partial replay \\
\hline
\end{tabular}
\end{center}
\caption{The 2 $\times$ 3 Whatever Recovery Landscape. The two-dimensions of Whatever recovery (accessibility of effects vs actions) and their implications on the classes of recovery algorithms possible}
\label{tab:recovery}
\end{table*}

Let us now see how we perform recovery and what levels of recovery are actually possible in the WLC model.

\subsection{Non-Consenting Organization Scenario}
\label{sec:non-consenting_scenario}

As described formally in Section~\ref{ssec:processing_model}, we must perform recovery on an organization~$O_l$ if $E_{l,t} \neq E_{cons,t}$. The reasons for this are multitude: It could be that~$O_l$ simply interpreted~$A_{l,t}$ differently than the others or that non-determinism is hidden in~$A_{l,t}$. It could also be that someone, e.g. an administrator, tampered with $E_{l,t}$.  

Irrespective of the cause, during recovery, we have to compute a new effect~$E'_{l,t}$. 
If $E'_{l,t} \neq E_{l,t}$, then the computed effect differs from the original effect, which was used for consensus, and $O_l$ has a chance to recover. 
If now $E'_{l,t} = E_{cons,t}$, then $O_l$ has recovered and can proceed with round~$t+1$. If not, then it can not recover and is excluded from the network. 

If no consensus has been reached at all, i.e. $E_{cons,t} = \textit{undefined}$, then we perform a new consensus round  on the new effects. The network can recover, if 
$$ \exists O'_{cons} = \{O_i, ..., O_j\}: |O'_{cons}| \ge c: E'_{i,t} = \ldots = E'_{j,t}\text{.}$$

\subsection{The 2 $\times$ 3 Recovery Landscape}

There are tradeoffs and optimizations involved in practical recovery. For example, instead of starting with an empty state and replaying the entire sequence of actions 
there are many more options. Table~\ref{tab:recovery} introduces the whatever recovery landscape. That landscape has two dimensions \textit{Effects} and \textit{Actions}. 
For the dimension \textit{Effects}, we distinguish between whether the \textit{state is not contained} and the \textit{state is contained}. For the dimension \textit{Actions}, we distinguish between \textit{not accessible}, \textit{blackbox actions} (we have access but do not understand the semantics in any way), and \textit{whitebox actions} (we have access \textbf{and} understand the semanics, i.e.~we see the individual operations carried out).
Like that we receive six different classes of recovery algorithms.
Note that each cell in that landscape includes all cells with weaker accessibility levels (i.e.~all cells that are further left and/or up). In the following we discuss each cell individually. We label each subsection with a visual notation of its position in the whatever recovery landscape, e.g.~\includegraphics[width=.03\textwidth]{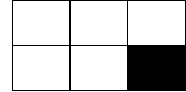} for (State: accessible, Actions: whitebox).

\subsection{No Recovery}

\noindent\textbf{Accessibility:}  \includegraphics[width=.03\textwidth]{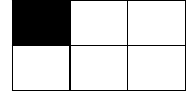} State: na, Actions: na. 

\noindent If neither state nor actions are available to us, we cannot recover a non-consenting organization.

\noindent\textbf{Recovery:} ---

\subsection{Recovery from a State}

\noindent\textbf{Accessibility:} \includegraphics[width=.03\textwidth]{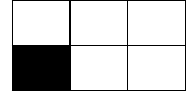} State: contained, Actions: na.

\noindent We have access to the state computed by the individual organizations but do not understand their semantics. 
As we don't have access to actions, we cannot apply them in any way.

\noindent\textbf{Recovery:} 
In order to recover organization~$O_l$ in round~$t$, we overwrite its local effect~$E_{l,t}$ with the effect~$E_{i,t}$ of \textit{any} other organization~$O_{i \neq l}$, that is matching the consensus effect~$E_{cons,t}$
$$E'_{l,t} = F_l(E_{i, t}, [])\text{.}$$
If~$E_{cons,t}=\text{undefined}$, i.e. no consensus was reached in round~$t$, then $O_l$ cannot recover.

\subsection{Full Replay}

\noindent\textbf{Accessibility:}  \includegraphics[width=.03\textwidth]{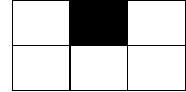}  State: na, Actions: blackbox.

\noindent The state is not contained, but we have access to the sequence of blackbox actions. 

\noindent\textbf{Recovery:} In order to recover organization~$O_l$ in round~$t$, we can (blindly) replay the entire history of blackbox actions~$A_{l,1},\ldots,A_{l,t}$ from the very beginning to restore the accumulated effect
$$E'_{l,t} = F_l(\emptyset, [A_{l,1},\ldots,A_{l,t}])\text{.}$$
If~$E'_{l,t}=E_{cons,t}$, then~$O_l$ may rejoin the network.

\subsection{Partial Replay from a State}

\noindent\textbf{Accessibility:} \includegraphics[width=.03\textwidth]{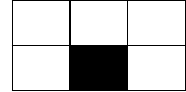}

\noindent State: contained, Actions: blackbox.

\noindent The state is contained. Further, we have access to blackbox actions but do not understand their semantics. 

\noindent\textbf{Recovery:} In order to recover~$O_l$ in round~$t$, we can perform a partial replay. In other words, we start with an older effect~$E_{l,s<t}$ and partially replay the blackbox actions~$A_{l,s+1},\ldots,A_{l,t}$ (\textit{redo} in database lingo):
$$E'_{l,t} = F_l(E_{l, s<t}, [A_{l,s+1},\ldots,A_{l,t}])$$
If~$E'_{l,t}=E_{cons,t}$, then~$O_l$ may rejoin the network.

\subsection{Optimized Full Replay}

\noindent\textbf{Accessibility:}  \includegraphics[width=.03\textwidth]{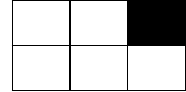} 

\noindent State: na, Actions: whitebox.

\noindent The effect does not contain the state, but we have access to the sequence of whitebox actions and understand the precise semantics of the actions. 

\noindent\textbf{Recovery:}  In order to recover organization~$O_l$ in round~$t$, we can replay the entire history of actions from the very beginning to restore the accumulated effect. 
However, as we have \textit{whitebox} actions available, we can also optimize the replay:
For example, if we know that an action~$A_{l,t}$ consists of a sequence of three transactions
$A_{l,t}=[T_1, T_2, T_3]\text{,}$ where all three transactions update the same record, then we could safely drop~$T_1$ and $T_2$ and use 
$A'_{l,t}=[T_3]$ for the replay.
Further possible optimizations include changing the order of operations and to allow for parallel execution of non-conflicting operations~\cite{parallel_exec}, as we show in Section~\ref{ssec:parallelism}. Thus, we restore the effect using the optimized actions~$A'_{l,1},\ldots,A'_{l,t}$:
$$E'_{l,t} = F_l(\emptyset, [A'_{l,1},\ldots,A'_{l,t}])$$
If~$E'_{l,t}=E_{cons,t}$, then~$O_l$ may rejoin the network.

\subsection{\mbox{Optimized Partial Replay from a State}}
\label{ssec:opt_partial_replay}

\noindent\textbf{Accessibility:}  \includegraphics[width=.03\textwidth]{recoverysymbols06}

\noindent State: contained, Actions: whitebox.

\noindent We have access to the state as well as access to whitebox actions.

\noindent\textbf{Recovery:} In order to recover~$O_l$ in round~$t$, we start with an older effect~$E_{l,s<t}$ and partially replay the optimized actions~$A'_{l,s+1},\ldots,A'_{l,t}$ (\textit{redo} in database lingo):
$$E'_{l,t} = F_l(E_{l, s<t}, [A'_{l,s+1},\ldots,A'_{l,t}])$$
If~$E'_{l,t}=E_{cons,t}$, then~$O_l$ may rejoin the network.

\subsection{Abstraction vs Implementation}

Notice that we use the concepts of `effect' and `action' to abstract from the details of a concrete implementation. 
For example, conceptually, it is not strictly necessary to physically materialize an effect: Replaying the entire history of actions is sufficient to restore an effect. This is similar to log-only databases (``the log is the database'')~\cite{octopus, logbase, logbase_indexing} that regard the database store as a performance optimized representation of the database log.

In ChainifyDB, which we will describe in detail in the next section, effects are materialized in form of database states and snapshots of those to enable high performance transaction processing and recovery. Actions represent blocks of transactions, which modify the database state and thereby generate new effects. Let us now see how it works in detail.

\section{ChainifyDB}
\label{sec:chainifydb}

In Section~\ref{sec:WLC} we have introduced and formalized our novel WLC model which is able to detect deviation of effects without making any assumptions on the behavior of the whatever-phase. In Section~\ref{sec:recovery}, we presented the different recovery options in the WLC landscape depending on the accessibility of effects and actions.

In this section we present ChainifyDB, a concrete system that instantiates the WLC model.
The core feature of ChainifyDB is to equip \textit{established} infrastructures, which already consist of several database management systems, with blockchain functionality as a layer on top. The challenge is that these infrastructures can be highly heterogeneous, i.e.~every participant potentially runs a \textit{different} DBMS where each system can interpret a certain transaction differently. As a result, the effects across participants might differ. 

As mentioned earlier, the classical OCE model, which relies on the previously discussed strong assumptions, is not capable of handling such a heterogeneous setup: The execution across participants is neither guaranteed to be deterministic nor equal. 
In contrast to that, our WLC model is perfectly suited to handle such heterogeneous scenarios, where no assumptions on the behavior of the organizations are made.

\subsection{Overview on our WLC-Implementation}
\label{sec:wlcimplementation}

Let us now see in detail how we implement the W-phase as well as the LC-phase in ChainifyDB. Just like our model, the implementation operates in rounds and consumes a batch of input transactions, which have been proposed by clients to the system, in every round. 

\begin{figure}[htb!]
    \begin{center}
  	\includegraphics[page=1, width=.7\linewidth, trim={0.5cm 15.6cm 7cm 0}, clip]{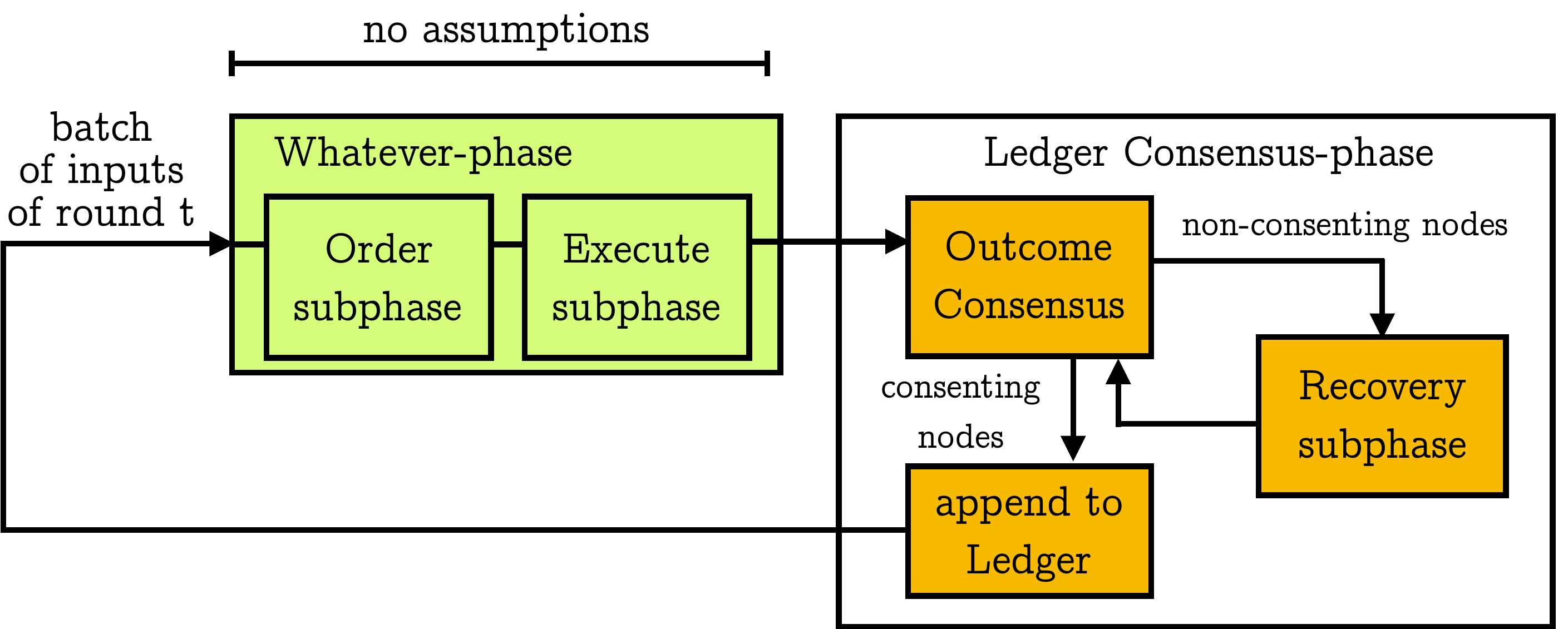}
    \end{center}
  \caption{ChainifyDB as a concrete instance of the Whatever-LedgerConsensus model (WLC).}
  \label{fig:wlc_chainify}
\end{figure}

\noindent\textbf{Whatever-phase}. 
In its simplest variant, ChainifyDB instantiates the W-phase of the WLC model with two subphases: the \textit{order-subphase} and the \textit{execute-subphase}. Figure~\ref{fig:wlc_chainify} visualizes the instantiation. 

\textit{Order-subphase}. In the order-subphase of round~$t$, a batch of input transaction is globally ordered and grouped in a block. Note that an action in our earlier formalization resembles a block of transactions here, i.e. $A_{l,t}=[T_1,T_2,T_3]$. Packing transactions into blocks is merely a performance optimization in order to amortize the costs for consensus later on. There is \textit{no} conceptual need to form blocks.
Notice that our order-subphase fully resembles the order-phase in the classical OCE-model. 
However, in strong contrast to the OCE-model, we do not perform a consensus round on the established order afterwards, even if we do not trust the ordering service in any way. For now, we simply take whatever the ordering service outputs.

\textit{Execute-subphase}. 
In the execute-subphase, each organization~$O_l$ receives an action~$A_{l,t}=[T_1, T_2, T_3]$ produced by the order-phase. Each transaction of that block, that has valid signatures, is then executed against the local relational database. This execution potentially updates the database and thereby produces an effect~$E_{l,t}$. In Section~\ref{ssec:digests} and Section~\ref{ssec:ledger}, we will outline in detail how this effect looks like in ChainifyDB. For now, let us simply assume that the effect captures all modifications done by the valid transactions of the block to the database.

Again, we want to point out that in contrast to the OCE-model, we do not assume deterministic execution across all organizations: The different DBMSs of two organizations could have interpreted a transaction slightly differently or two organizations could have received different blocks from the order-phase altogether.

\noindent \textbf{LedgerConsensus-phase}.
In the ledger consensus phase, all organizations have to reach consensus on the effects produced by the whatever-phase of the individual organizations. 
Thus, in a consensus round, which we will describe in Section~\ref{ssec:consensus} in detail, the organizations first try to agree on one particular effect. Then, each organization whose effect is consenting commits it to its local ledger and proceeds with round~$t+1$. 
Again, only if consensus on an effect is reached, we consider it as globally committed. 

\textit{Recovery-subphase}. 
If the effect of an organization is non-consenting, the organization must at least try to recover from this situation. This is done using a variant of \textit{Optimized Partial Replay from a (Logical) Snapshot}
 \includegraphics[width=.02\textwidth]{recoverysymbols06}   
as introduced in Section~\ref{sec:recovery}. 
We will explain our recovery mechanism in detail in Section~\ref{ssec:recovery_phase}.

\subsection{Logical per Block Digests}
\label{ssec:digests}

In the previous section, we mentioned that the execution of a block on the local database produces an effect as a side-product of execution. On this effect, the consensus round is performed. It is also eventually committed to the ledger.  
Thus, let us see in the following how we precisely generate the effect.

\begin{figure}[htb!]
  \begin{center}
  \includegraphics[page=1, width=.7\linewidth, trim={1cm 26cm 62cm 0}, clip]{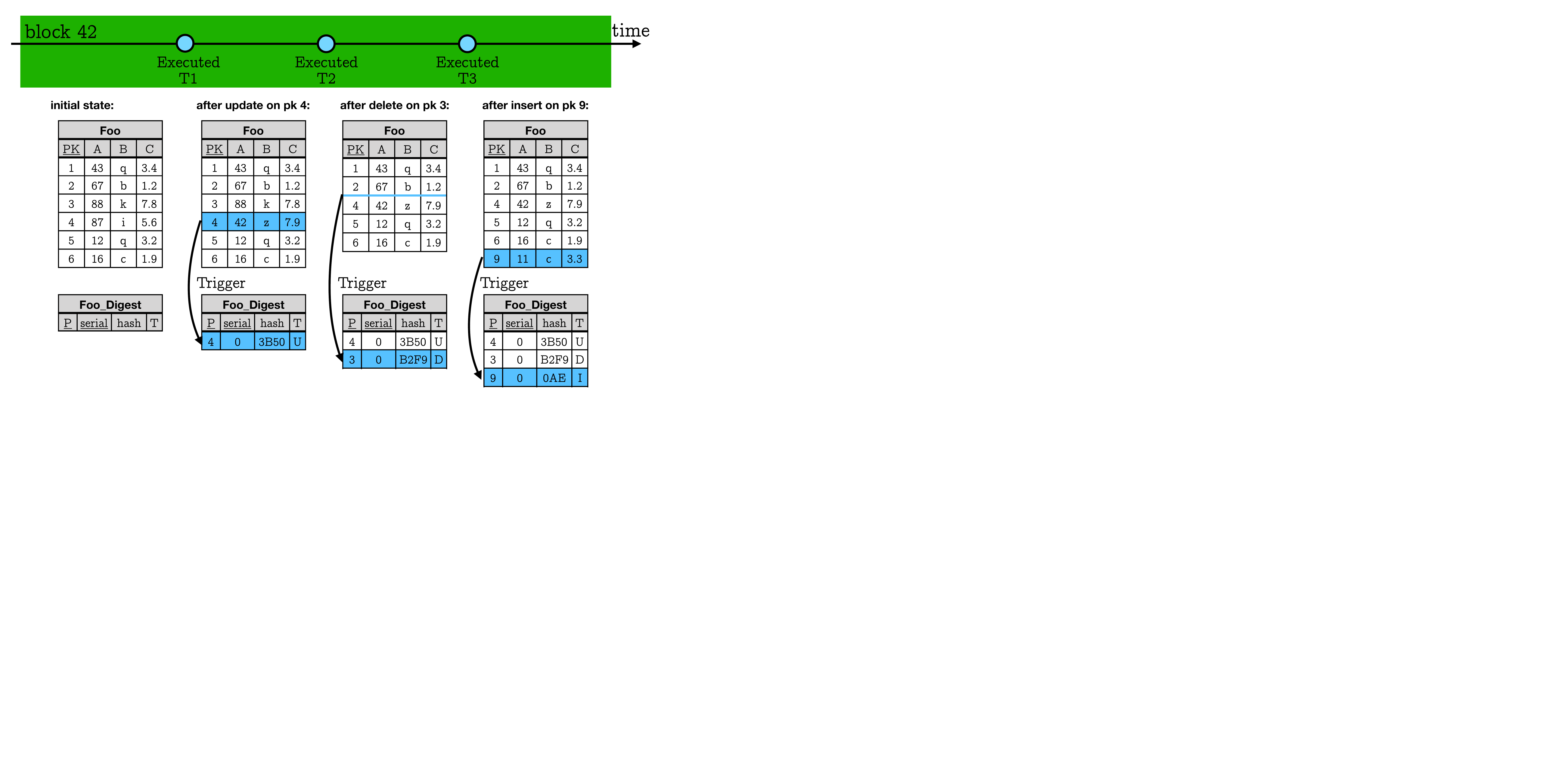}
  \end{center}
  \caption{Logical tuple-wise per block digest computation on an example table Foo. All changes are automatically tracked and digested through SQL 99 triggers.}
  \label{fig:difftables}
\end{figure}

In ChainifyDB we assume SQL-99 compliant relational DBMSs to keep the state at each organization. This has two reasons: 
On the one hand, we want to allow for existing organizations with existing DBMS-products to be able to easily build WLC-networks with blockchain-style guarantees. 
On the other hand, we can utilize SQL 99 triggers to realize a vendor-independent digest versioning mechanism, that specifically versions the data of ChainifyDB in form of a \textit{digest table}. 

The digest table is computed per block. We instrument every shared table in our system with a trigger mechanism to automatically populate this digest as follows: for every tuple changed by an \smalltt{INSERT}, \smalltt{UPDATE}, and \smalltt{DELETE}-statement, we create a corresponding \textit{digest tuple}. A digest tuple has the following schema:
\smalltt{[PK:<as of Foo>, serial:int, hash:int, T]}.
Here, \smalltt{PK} is the primary key of the original table (which may of course also be a compound key), \smalltt{serial} is a strictly monotonously rising counter used to distinguish entries with the same \smalltt{PK} (every new version of a tuple increases this counter), \smalltt{hash} is the digest of the values of the tuple after it was changed (for a delete: its state before removing it from the table), \smalltt{T} is the type of change that was performed, i.e.~(I)nsert, (U)pdate or (D)elete. 
Notice that in contrast to recovery algorithms like ARIES~\cite{aries}, in those tuples we do not preserve the information how to undo/redo changes, as we simply do not need that information. 

Figure \ref{fig:difftables} shows an example how to process a block of three transactions: we start with a particular state of table \smalltt{Foo} and an empty digest table \smalltt{Foo\_Digest}. Now, we perform an update on tuple with \smalltt{PK}=4. As a result, the tuple in \smalltt{Foo} is changed to (4,~42,~z,~7.9) and we insert a new digest tuple (4,~0,~3B50,~U) into \smalltt{Foo\_Digest}. 
After that a delete of record with PK=3 is performed. 
The tuple in \smalltt{Foo} is deleted and we insert a new digest tuple (3,~0,~B2F9,~D) into \smalltt{Foo\_Digest}. We proceed until we processed all three transactions TA1--TA3 available in this particular block. For the next block to process we start with an empty digest table.

\subsection{LedgerBlocks}
\label{ssec:ledger}

Although the digest table captures all changes done to a table by the last block of transactions, it does not represent the effect yet. The actual effect is represented in form of a so called \smalltt{LedgerBlock}, which consists of the following fields:

\begin{enumerate}
\item \smalltt{TA\_list}: The list of transactions that were part of this block. Each transaction is represented by its SQL code.
\item \smalltt{TA\_successful}: A bitlist flagging the successfully executed transactions. This is important as transactions may of course fail and that behavior should be part of consensus across all organizations.
\item \smalltt{hash\_digest}: A hash of the logical contents of the digest table. In our case, this is a hash over the hash-values present in the diff table. The hash values are concatenated in lexicographical [PK, serial]-order and then input into SHA256.
\item \smalltt{hash\_previous\_LedgerBlock}: A hash value of the entire contents of the previous \smalltt{LedgerBlock} appended to the ledger, again in form of a SHA256 hash. This backward chaining of hashes allows anyone to verify the integrity of the ledger in linear time.
\end{enumerate}

This \smalltt{LedgerBlock} now leaves the W-phase and enters the LC-phase to determine whether consensus can be reached. 

\subsection{Consensus Algorithm}
\label{ssec:consensus}

In our permissioned setup, we can safely make the assumption, that all organizations of the network are known at any time and that no organization can join the network during a consensus round. This allows us to use a lightweight voting algorithm for this purpose, instead of having to rely on more heavyweight consensus algorithms such as~\cite{pbft, optimistic_bft, parallel_bft, efficient_bft}. To determine whether consensus was reached, a \textit{consensus policy}~$c$ must be specified by all organizations in advance during the bootstrapping process of the network. The constant~$c$ specifies how many organizations must have reached the same effect. 

Our algorithm is shown in Algorithm~\ref{alg:consensus} in the appendix and resembles the perspective of a single organization.

In the first step of the consensus algorithm, the individual organization has to count how often each \smalltt{LedgerBlock} occurred in the network. To do so, it requests the so called \smalltt{LedgerBlockHashes} from all other organizations and counts the occurrences, including its own local \smalltt{LedgerBlockHash}. This \smalltt{LedgerBlockHash} is essentially a compressed form of the contents of a \smalltt{LedgerBlock}.

Consensus is reached if two conditions hold: (a)~There must be a \smalltt{LedgerBlockHash} that occurred at least $c$~times. (b)~This consensus \smalltt{LedgerBlockHash} equals the local \smalltt{LedgerBlockHash}.
If both hold, then the organization can append/commit its \smalltt{LedgerBlock} to its local ledger.

\subsection{Logical Checkpointing and Recovery}
\label{ssec:recovery_phase}

If an organization is non-consenting, it must enter recovery as outlined in Section~\ref{sec:wlcimplementation} and presented in Figure~\ref{fig:wlc_chainify}. 

\begin{figure}[htb!]
	\includegraphics[page=1, width=\linewidth, trim={3cm 27cm 78cm 0}, clip]{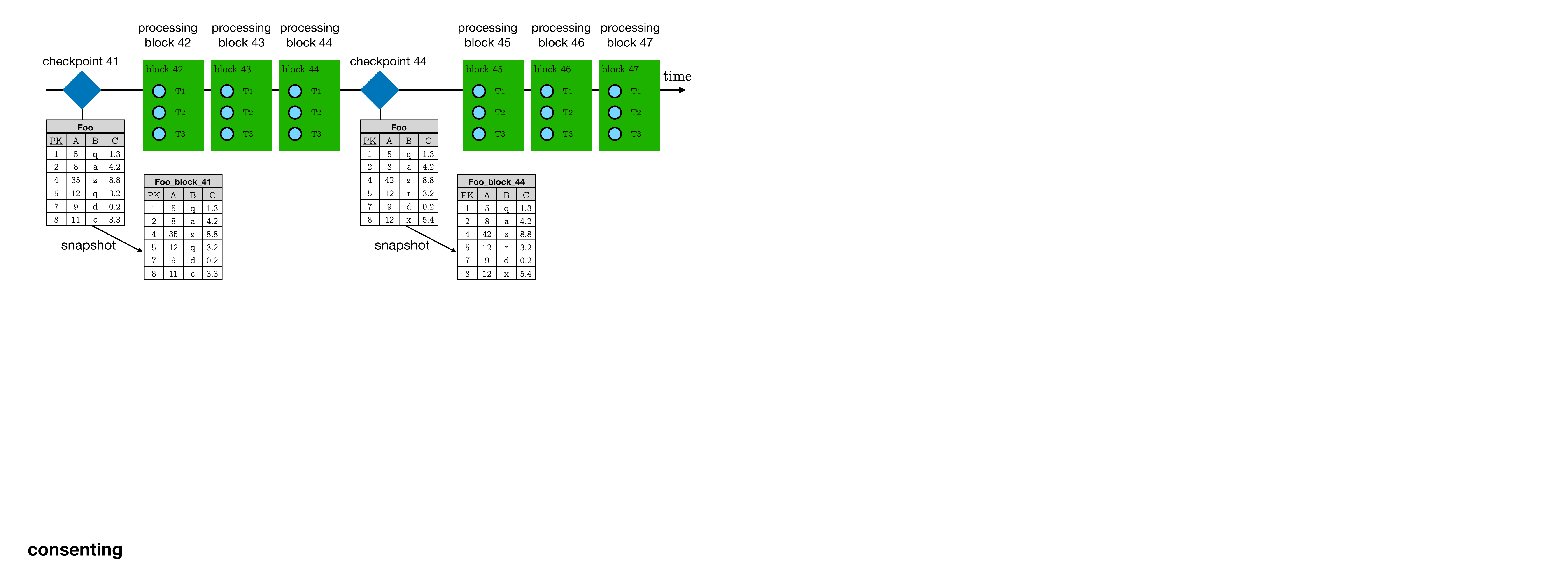}  \caption{ChainifyDB's checkpointing mechanism. Here, a checkpoint is created after every three blocks. }
\label{fig:recovery:snapshot}
\end{figure}

\begin{figure*}[htb!]
	\includegraphics[page=2, width=\linewidth, trim={0 23cm 10cm 0}, clip]{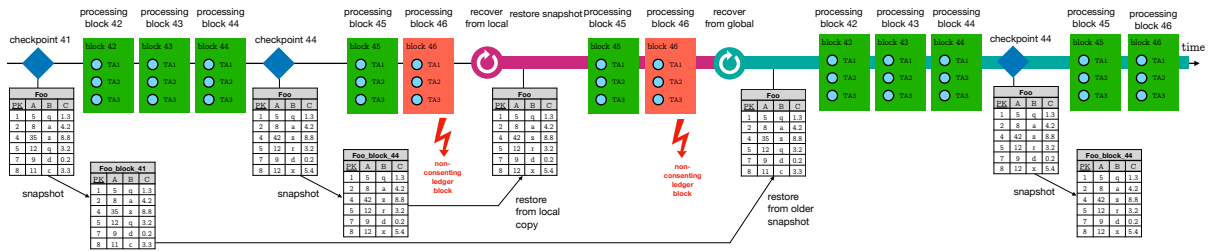}  \caption{ChainifyDB's Recovery using checkpoints. As block 46 is non-consenting it has to enter the recovery phase. It will first try to recover using the most recent local checkpoint. This fails in this example and hence recovery from an older checkpoint is performed.}
  \label{fig:recovery:recover}
\end{figure*}

For recovery, ChainifyDB implements \textit{Optimized Partial Replay from a (Logical) Snapshot}
 \includegraphics[width=.02\textwidth]{recoverysymbols06}   
as introduced in Section~\ref{ssec:opt_partial_replay}.  Again, like the digests (see Section~\ref{ssec:digests}) our approach is \textit{DBMS-system independent}: we do not need access to the source code of the DBMS.
The Figures~\ref{fig:recovery:snapshot}~and~\ref{fig:recovery:recover} show an example run of our checkpointing and recovery algorithm.

Figure~\ref{fig:recovery:snapshot} shows the normal operation mode of a single organization that is consenting: we create a checkpoint by creating a snapshot of table \smalltt{Foo} after every $k$ blocks ($k=3$ in the example). Snapshots are created on the SQL-level through either a non-maintained materialized view or by a \smalltt{CREATE TABLE} command. If the source code of the DBMS and the operating system is available, the snapshotting support from the operating system could be exploited at this point as well~\cite{DBLP:conf/sigmod/SharmaSD18, hyper} -- however, we do not make this assumption in our design. The snapshot is created for all tables that were changed since the last consistent snapshot. Creating such a snapshot is surprisingly fast: Snapshotting the accounts table with $1{,}000{,}000$~rows, which is part of the Smallbank~\cite{smallbank} benchmark used in our experiment evaluation, takes only $827$ms in total on our machine of type~$2$ running PostgreSQL (see Section~\ref{sec:eval}). 
Notice that there is no need to store this checkpoint in the ledger, as done in ARIES~\cite{aries} for instance: As the information contained in a checkpoint can be fully recomputed from the ledger, it has to pass the LC-phase anyways again.

Figure~\ref{fig:recovery:recover} shows an organization switching to recovery mode. This organization is in normal (consenting) mode for all blocks shown up to and including block~45. For block~46 this organization is non-consenting. Hence, it must enter the recovery phase. 

First, we have to reset \smalltt{Foo} to the state of the latest consistent snapshot \smalltt{Foo\_block\_44}. Then, we replay block~45 which is consenting. Then, we replay block~46 which is unfortunately \textit{again} non-consenting. In this situation we have to assume that this local snapshot has an issue, i.e. it was corrupted externally. Thus, we reset the local table \smalltt{Foo} to the second latest snapshot \smalltt{Foo\_block\_41}. Now, we replay all blocks starting from block~42 up to block~46. This time block~46 is consenting. 

In our implementation, we keep three committed snapshots per organizations. If replaying from all of these snapshots does not lead to a consenting organization, then we can try to replay the \textit{entire} history. If even this fails, we have to assume that the ordering service acts maliciously and can try again after starting a fresh ordering service, possibly by a different organization.  

Notice that, of course, there is no 100\% guarantee that any of these measures will lead to a consenting organization. Severe problems such as a hardware error (in particular an error that is not detected and transparently fixed by hardware or operating system itself) are out of reach for repair by ChainifyDB.  
The important point here is that we detect the problem early on (after every block of transactions and not only eventually), try to ``rehabilitate''  this organization through recovery, and reintegrate it into the network. 

Our recovery algorithm is summarized in Algorithm~\ref{alg:recovery} in the appendix.

\section{Optimizations}
\label{sec:optimizations}

Before we come to the experimental evaluation of ChainifyDB, let us discuss a set of interesting optimizations.

\subsection{Transaction Agreement}
\label{ssec:agreement}

A powerful feature of ChainifyDB is that clients can propose arbitrary transactions to the system. These transactions are then simply executed against the local DBMSs of the individual organizations without any restrictions. 
However, there might be situations where such a plain execution without any restrictions is highly undesired by the organizations. 

For example, consider the scenario where two organizations would like to log their mutual trades in a shared table. A transaction inserting a trade could look as follows

\noindent
\begin{minipage}{\linewidth}
\begin{footnotesize}
\begin{lstlisting}[numbers=none]
INSERT INTO Trades(TID,product,amount,totalprice) 
VALUES (42,"Gearbox",5,60000):
\end{lstlisting}
\end{footnotesize}
\end{minipage}

\noindent Obviously, this transaction is only meaningful if certain integrity constraints hold: The selling organization must have enough products in stock (at least five gearboxes in our example) and the buying organization must have enough money available (at least $60{,}000$ in our example). 
To enforce such integrity constraints, we prepend an optional \textit{agreement-subphase} to our pipeline, through which any transaction proposed to ChainifyDB must go first. Only if all involved organizations agree to the proposed transaction, the transaction may enter the subsequent order-phase.

To enable the optional agreement phase, two steps must be carried out by the organizations: First, an \textit{agreement policy} must be installed in consent when creating the shared table. It specifies for the shared table which organizations have to agree upon a transaction operating on that table. For our trading example, the policy of the table \textit{Trades} could look as follows, enforcing both involved organizations to decide for agreement:

\noindent
\begin{minipage}{\linewidth}
\begin{footnotesize}
\begin{lstlisting}[numbers=none]
AgreementPolicy(Trades) = { SellingOrganization, BuyingOrganization }
\end{lstlisting}
\end{footnotesize}
\end{minipage}

\noindent Second, each organization has to implement its individual integrity constraints, which are evaluated against each proposed transaction. Note that these constraints could be formulated to compare the transaction against local data.  
For example, the agreements for the selling organization and the buying organization could look as follows:

\noindent
\begin{minipage}{\linewidth}
\begin{footnotesize}
\begin{lstlisting}[numbers=none]
SellingOrganization.agree(T) { 
   return T.amount <= Stocks.amount WHERE T.product == Stocks.product 
}

BuyingOrganization.agree(T) = { 
   return T.totalprice <= Fund.availableMoney 
}
\end{lstlisting}
\end{footnotesize}
\end{minipage}

Only if both organizations agree to a transaction operating on the \textit{Trades} table, it is passed on for further processing.

\subsection{Iterative WLC-Setups}
\label{ssec:iterative}

\begin{figure}[htb!]
  \includegraphics[page=2, width=1.2\linewidth, trim={0 6.5cm 3 0}, clip]{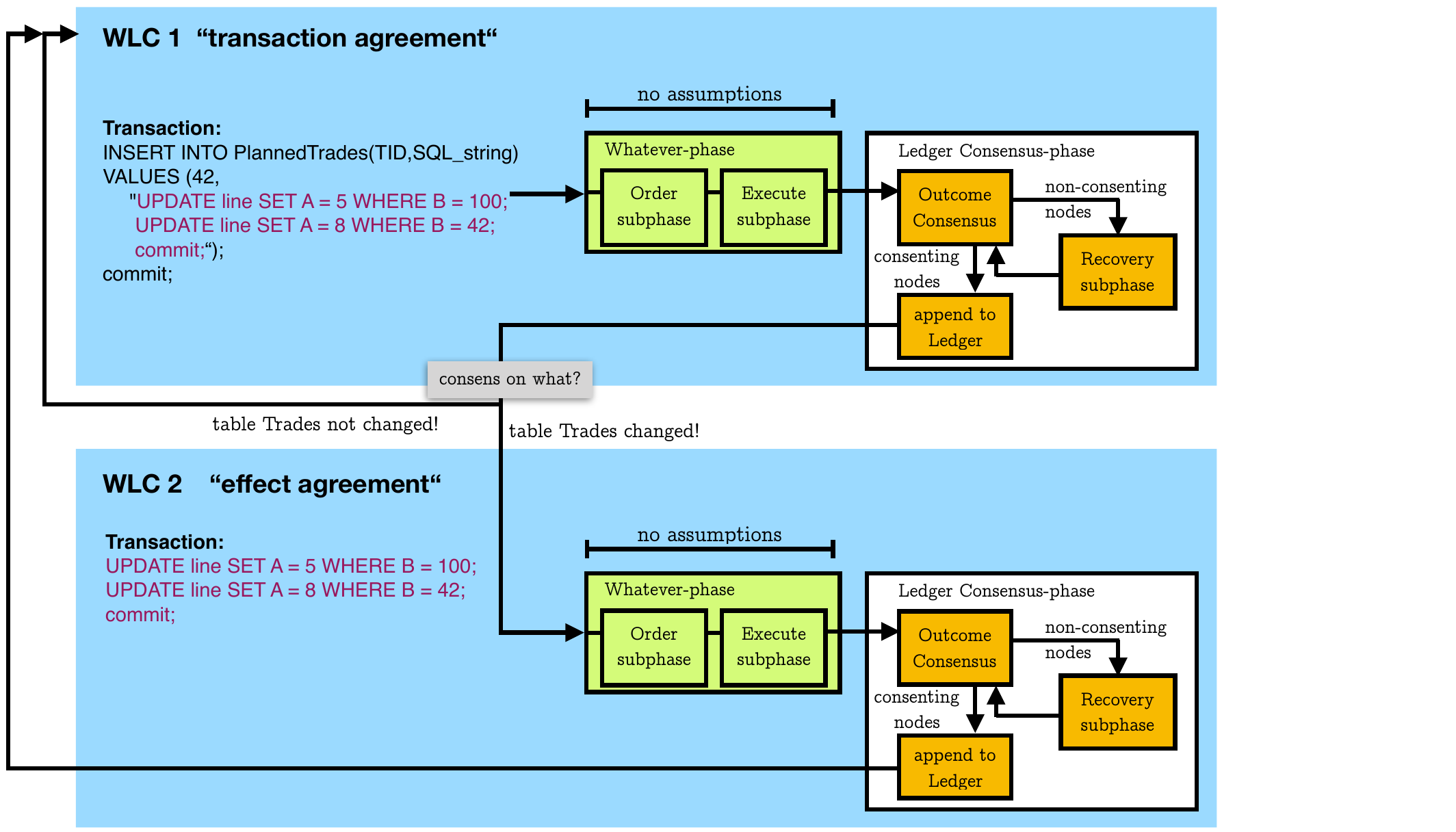}
  \caption{Transaction agreement can be regarded as running a separate pre-WLC phase on transaction agreement before executing the actual transaction.}
  \label{fig:wlc_iterative}
\end{figure}

From a 10,000 feet perspective an agreement can be considered a different WLC-iteration where the participants agree upon the string describing the SQL-transaction to be executed rather than the outcome of running that transaction. This is visualized in Figure~\ref{fig:wlc_iterative}. In WLC~1 the organizations must first agree upon a trade to be done (represented as an SQL-transaction). If the organizations agree, a SQL string is inserted into a  table \smalltt{PlannedTrades}:

\noindent
\begin{minipage}{\linewidth}
\begin{footnotesize}
\begin{lstlisting}
INSERT INTO PlannedTrades(TID,SQL_string) 
VALUES (42,"UPDATE line SET..."):
\end{lstlisting}
\end{footnotesize}
\end{minipage}
If there is consent, that the tuple was inserted, i.e.~that this trade should be done, the corresponding SQL string will be send to WLC~2 which executes the SQL string.

\noindent In summary, mapping subphases to incremental rounds of the WLC model enables interesting abstractions. In future work, we plan to investigate these mappings in depth. 

\subsection{Parallel Transaction Execution}
\label{ssec:parallelism}

Apart from the interplay of the phases, we did not precisely specify how the execute-subphase actually runs transactions in the underlying database system. 

Naively, we could simply execute all valid transactions of a block one by one in a sequential fashion. However, this strategy drastically wastes performance if the underlying system is able to execute transactions in parallel. 

This leads us to an alternative strategy, where we could simply submit all valid transactions of a block to the underlying (potentially parallel) database system in one batch and let it execute them concurrently. While this strategy leverages the performance of the underlying system, it creates another problem: it is very likely that every DBMS schedules the same batch of transactions differently for parallel execution. As a consequence, the commit order of the transactions likely differs across organizations, thus increasing the likelihood of non-consent.  

The strategy we apply in ChainifyDB sits right between the previously mentioned strategies and is inspired by the parallel transaction execution proposed in~\cite{parallel_exec} and relates to the ideas of~\cite{early_write_visibility, batching, quecc}. When a block of transactions is received by the execute-subphase, we first identify all existing conflict dependencies between transactions. This allows us to form mini-batches of transactions, that can be executed safely in parallel, as they have no conflicting dependencies. 

Let us see in detail how it works. The process can be decomposed into three phases:

\noindent (1)~\textbf{Semantic Analysis}.~First, for a block of transactions, we do a semantic analysis of each transaction. Effectively, this means parsing the SQL statements and extracting the read and write set of each transaction. These read and write sets are realized as intervals on the accessed columns to support capturing both point query and range query accesses. 
For instance, assume the following two single-statement transactions:

\noindent
\begin{minipage}{\linewidth}
\begin{footnotesize}
\begin{lstlisting}[numbers=none]
   T1:UPDATE Foo SET A = 5 WHERE PK = 100;
   T2:UPDATE Foo SET A = 8 WHERE PK > 42;
\end{lstlisting}
\end{footnotesize}
\end{minipage}

\noindent The extracted intervals for these transactions are:

\noindent
\begin{minipage}{\linewidth}
\begin{footnotesize}
\begin{lstlisting}[numbers=none]
   T1: A is updated where PK is in [100,100] 
   T2: A is updated where PK is in [42,infinity]
\end{lstlisting}
\end{footnotesize}
\end{minipage}

\noindent (2)~\textbf{Creating the Dependency Graph}.~With the intervals at hand, we can create the dependency graph for the block. For two transactions having a read-write, write-write, or write-read conflict, we add a corresponding edge to the graph. Note that as transactions are inserted into the dependency graph in the execution order given by the block, no cycles can occur in the graph. 

\begin{figure}[htb!]
\begin{center}
  \includegraphics[page=1, width=.4\linewidth, trim={10cm 18cm 14cm 1.8cm}, clip]{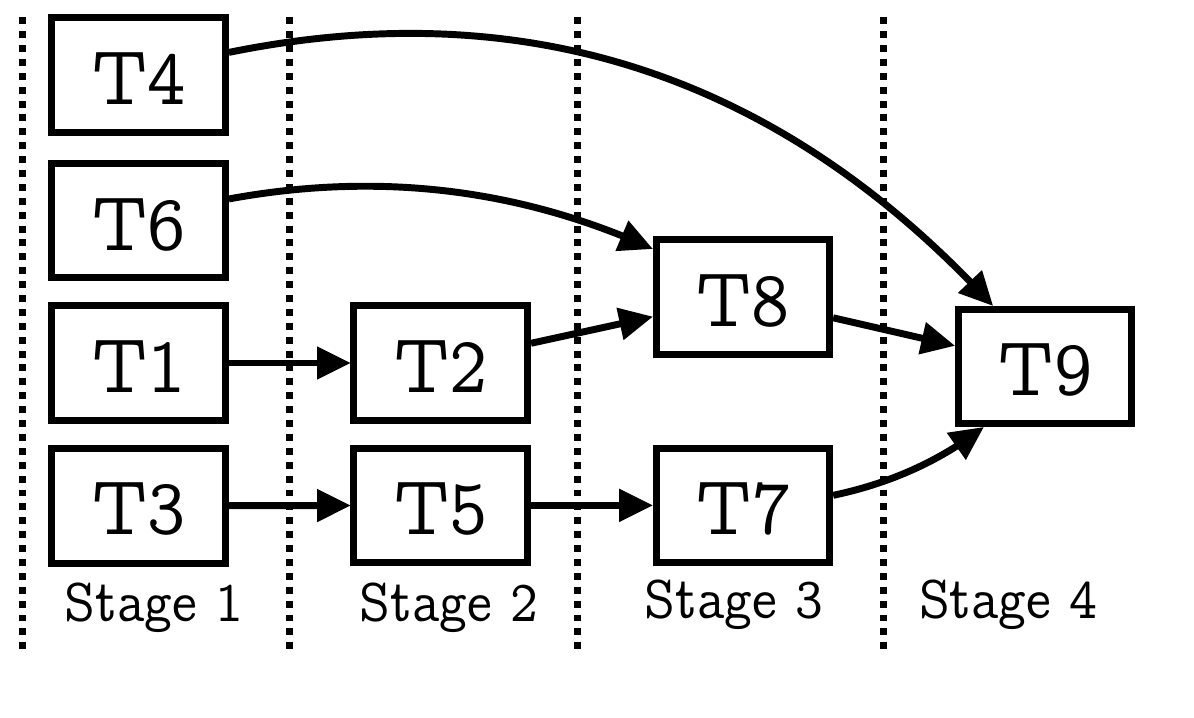}
  \end{center}
  \caption{A topological sort of the dependency graph with $k=9$ transactions yielding four execution stages.}
  \label{fig:graph}
\end{figure}

Let us extend the example from our Semantic Analysis Phase and let us assume, that T1 has been added to the dependency graph already. By inspecting T2 we can determine that PK[42, inf] overlaps with PK[100,100] of T1. As T2 is an update transaction, there is a conflict between T2 and T1 and add a dependency edge from T1 to T2.
Figure~\ref{fig:graph} presents an example dependency graph for $9$~transactions.  

\noindent (3)~\textbf{Executing the Dependency Graph}.~Finally, we can start executing the transactions in parallel. To do so, we perform topological sorting, i.e.~we traverse the execution stages of the graph, that are implicitly given by the dependencies. Our example graph in Figure~\ref{fig:graph} has four stages in total. Within one stage, all transactions can be executed in parallel, as no dependencies exist between those transactions.

The actual parallel execution on the underlying database system is realized using $k$~client connections to the DBMS. To execute the transactions within an execution stage in parallel, the $k$~clients pick transactions from the stage and submit them to the underlying system. As our method is conflict free, it guarantees the generation of serializable schedules.

Therefore we can basically switch off concurrency control on the underlying system. This can be done by setting the isolation level of the underlying system to READ~UNCOMMITTED\footnote{Note however, that typical MVCC-implementations do not provide this level, e.g.~in PostgreSQL the weakest isolation level possible is READ COMMITTED.}~\cite{isolation_levels} to get the best performance out of the DBMS.

\section{System Architecture}
\label{sec:arch}

With a good knowledge of the logical design of ChainifyDB in Section~\ref{sec:chainifydb}, let us now map this logical design to the physical architecture of ChainifyDB. A ChainifyDB network consists of $n$~($\ge3$) organizations, an untrusted OrderingServer, and a set of Kafka nodes that act as a message broker to distribute the block created by the OrderingServer. The ChainifyDB system is implemented using \smalltt{Golang} and \smalltt{C++} as a primary language in just under 11,000 LOC, excluding the dependencies and the generated code.

Each participating organization runs one or more instances of ChainifyServer, AgreementServer, Execution- Server, and the ConsensusServer. Unless stated otherwise, we use the SHA256 hashing algorithm from the \smalltt{crypto} package, and the Sign and Verify algorithms from the \smalltt{elliptic} package of Golang 1.10.4. All messages exchanged between different components of the ChainifyDB network are signed to preserve the integrity of the messages flowing in the system. And unless stated otherwise, we use \smalltt{RPC} calls~\cite{rpc, grpc} to communicate between different components of the system.

\noindent \textbf{ChainifyServer}: This component receives the signed transaction in the form of a proposal from the client. Each ChainifyServer stores the policy that defines the set of organizations that must agree to this transaction. Using this policy, the ChainifyServer communicates with the AgreementServer of the responsible organizations to collect all the agreements required for this transaction. It then packs the original proposal and all the agreements into a \smalltt{ChainedTransaction}. If all agreements are valid, it forwards the ChainedTransaction to the OrderingServer. Each organization can run multiple instances of ChainifyServer to linearly scale the number of client requests handled by this organization.

\noindent \textbf{AgreementServer}: This component is responsible for validating the local integrity constraints of the organization. It receives the transaction sent by the ChainifyServer using the RPC call, decides whether this transaction passes the local integrity check, and responds with a signed \smalltt{yes} or \smalltt{no}. The organization can run multiple instances of AgreementServer to linearly scale the agreement-phase as well.

\noindent \textbf{OrderingServer}: This component receives the ChainedTransactions from the ChainifyServer and packs them into a block in FIFO order. If the block has a sufficient size or a certain amount of time has elapsed, the block is cut and forwarded to Apache Kafka The Kafka service then delivers the block to all ExecutionServers. This component can be also scaled linearly to accommodate the high frequency of incoming chained transactions.

\noindent \textbf{ExecutionServer}: This component fetches the blocks created by the OrderingServer and executes them in-order on the underlying database system. The ExecutionServer first verifies all the signatures to make sure that no one in the transaction pipeline tampered the transaction.  If the signatures and the agreements are valid, the transaction is marked as valid and invalid otherwise. The ExecutionServer then forwards the block to the \smalltt{GraphGenerator} implemented in \smalltt{C++}, which generates an efficient execution graph for this block as described in Section~\ref{ssec:parallelism}. After receiving the optimized execution-graph,  it is executed on the database system in parallel. During the execution, ChainifyDB collects the digest of the execution using SQL 99 triggers. 
After the execution of all transactions in the block, the ExecutionServer generates the \smalltt{LedgerBlock} and the corresponding \smalltt{LedgerBlockHash} as described in Section~\ref{ssec:ledger}, and forwards the \smalltt{LedgerBlockHash} to the ConsensusServer for the consensus round.
The organization can run multiple instances of this component to efficiently replicate the effects of the transactions on different database systems inside the organization.

\noindent \textbf{ConsensusServer}: This component receives the \smalltt{Ledger Block} and the \smalltt{LedgerBlockHash} from the ExecutionServer. It then communicates with other organization's ConsensusServer to check whether a particular \smalltt{LedgerBlockHash} reaches consensus. On reaching consensus, the ConsensusServer appends the corresponding \smalltt{LedgerBlock} to the ledger. Otherwise, the organization performs recovery.

In summary, all ChainifyDB components up to the ConsensusServer fall into the W-phase, while the ConsensusServer itself falls into to the LC-phase. Note that if we purely follow the WLC model, the transactions of the block could be excluded from the consensus round and from the ledger, since technically, we do not care about \textit{how} an organization reaches a state. However, we still keep the information about the transactions per block to enable a more sophisticated recovery phase.

\section{Experimental Evaluation}
\label{sec:eval}

To evaluate ChainifyDB we use the following system setup.

\subsection{Setup and Workload}
\label{ssec:setup}

\noindent \textbf{Type 1 (small)}: Two quad-core Intel Xeon CPU E5-2407 running at $2.2$~GHz, equipped with $48$GB of DDR3 RAM.

\noindent \textbf{Type 2 (large)}: Two hexa-core Intel Xeon CPU X5690 running at $3.47$~GHz, equipped with $192$GB of DDR3 RAM.

Unless stated otherwise, we use a heterogeneous network consisting of three independent organizations~$O_1$, $O_2$, and $O_3$. Organization~$O_1$ owns two machines of type~1, where PostgreSQL~$11.2$ is running on one of these machines.  Organization~$O_2$ owns two machines of type~1 as well, but MySQL~$8.0.18$ is running on one of them. Finally, organization~$O_3$ owns two machines of type~2, where again PostgreSQL is running on one of the machines. The individual components of ChainifyDB, as described in Section~\ref{sec:arch}, are automatically distributed across the two machines of each organization. Additionally, there is a dedicated machine of type~2 that represents the client firing transactions to ChainifyDB as well as a type~2 machine that solely runs the ordering service.

As consensus policy, we configure two different options: In the first option \textit{Any-2} we set $c=2$ such that at least two out of our three organizations have to produce the same effect to reach consensus. In the second option~\textit{All-3} we set $c=3$ and consensus is reached only if all three organizations produce the same effect. 
In any case, all three organizations have to agree to every transaction. 
Besides, empirical evaluation revealed a block size of $4096$~transactions to be a good fit (see Section~\ref{ssec:blocksize}). Of course, we also activate parallel transaction execution as described in Section~\ref{ssec:parallelism}.

As workload we use transactions from Smallbank~\cite{smallbank}, which simulate a typical asset transfer scenario. To bootstrap the test, we create for $100{,}000$~users a checking account and a savings account each and initialize them with random balances. 
The workload consists of the following four transactions: \textit{TransactSavings} and \textit{DepositChecking} increase the savings account and the checking account by a certain amount. \textit{SendPayment} transfers money between two given checking accounts. \textit{WriteCheck} decreases a checking account by a certain amount.
During a single run, we repeatedly fire these four transactions at a fire rate of $4096$~transactions per client, where we uniformly pick one of the transactions in a random fashion. For each picked transaction, we determine the accounts to access based on a Zipfian distribution with a $s$-value of~$1.1$ and a $v$-value of~$1$, unless stated otherwise.

\subsection{Throughput}
\label{ssec:throughput}
\vspace*{-0.1cm}

\begin{figure}[!htb]
\begin{center}
\subfigure[Throughput of ChainifyDB with \textit{Any-2} and \textit{All-3} policy for varying number of clients. Additionally, we evaluate Fabric~\cite{fabric} and Fabric++~\cite{blurring_the_lines}. We use the Smallbank workload following a Zipf distribution.]{
\includegraphics[width=.7\linewidth, page=1, trim={2cm 2cm 2cm 5cm}, clip]{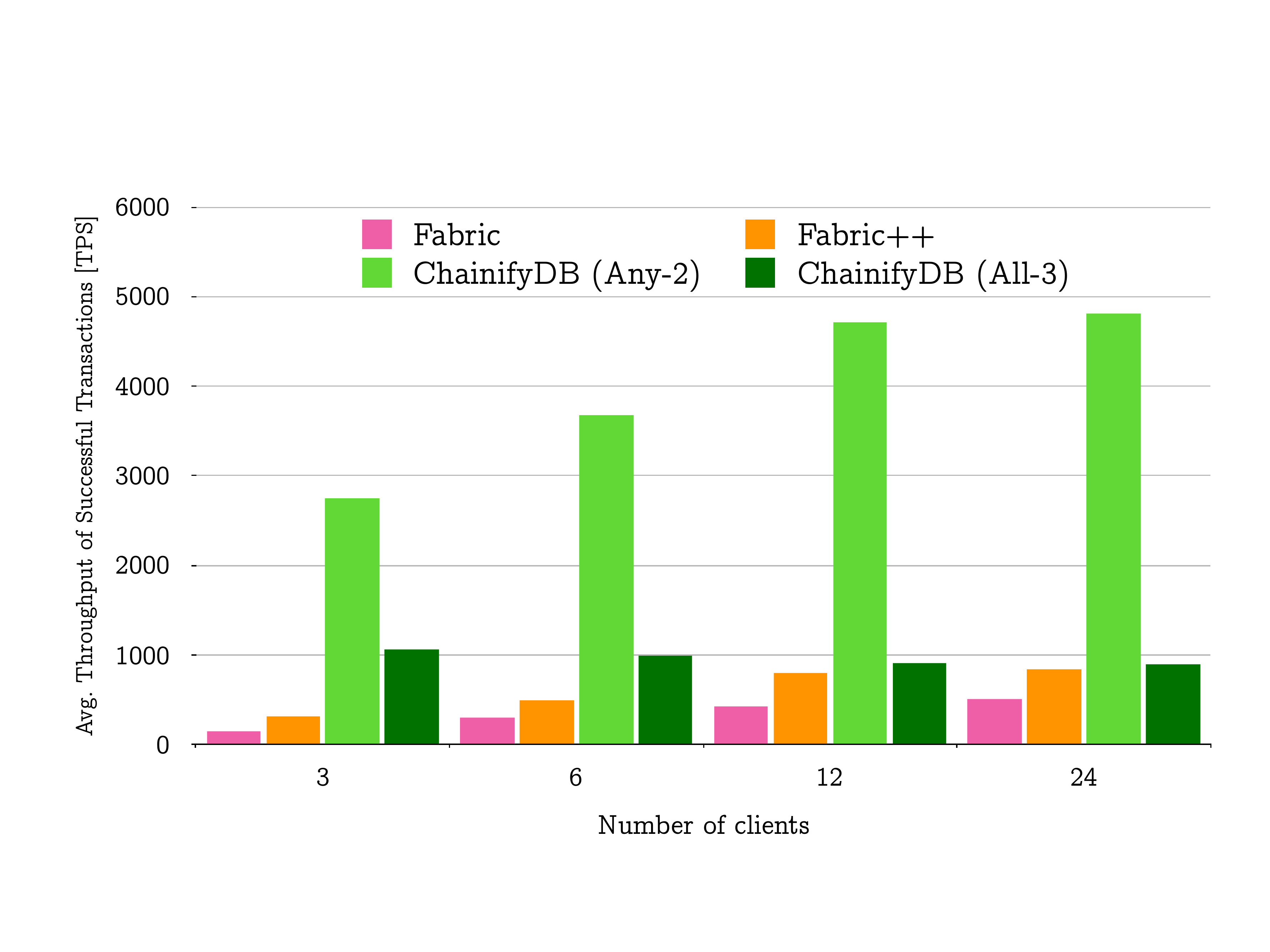}
\label{figs:throughput1}
}
\subfigure[Throughput of standalone MySQL and PostgreSQL for varying number of clients. The same workload as in Figure~\ref{figs:throughput1} is fired using OLTP-bench. Note that OLTP-bench follows a uniform distribution.]{
\includegraphics[width=.7\linewidth, page=2, trim={2cm 2cm 2cm 5cm}, clip]{plots.pdf}
\label{figs:throughput2}
}
\end{center}
\caption{Throughput of successful transactions for the heterogeneous setup as described in Section~\ref{ssec:setup}.}
\label{figs:throughput}
\end{figure}

We start the experimental evaluation of ChainifyDB by inspecting the most important metric of a blockchain system: the throughput of successful transactions, that make it through the system.  

Therefore, we first inspect the throughput of ChainifyDB in our heterogeneous setup under our two different consensus policies~Any-2 and All-3. 
Additionally to ChainifyDB, we show the following two PBS~baselines: (a)~Vanilla Fabric~\cite{fabric}~v1.2, probably the most prominent PBS system currently. (b)~Fabric++~\cite{blurring_the_lines}, an improved version of Fabric~v1.2. Both Fabric and Fabric++ are also distributed across the same set of machines and the blocksize is set to~$1024$.

Figure~\ref{figs:throughput1} shows the results. On the $x$-axis, we vary the number of clients firing transactions concurrently from $3$~clients to $24$~clients. On the $y$-axis, we show the average throughput of successful transactions, excluding a ramp-up phase of the first five seconds. 
We can observe that ChainifyDB using the Any-2 strategy shows a significantly higher throughput than Fabric++ with up to almost $5000$~transactions per second. In comparison, Fabric++ achieves only around~$1000$~transactions per second, although it makes considerably more assumptions than ChainifyDB: First, it assumes the ordering service to be trustworthy. Second, it assumes the execution to be deterministic and therefore does not perform any consensus round on the state.  

Regarding ChainifyDB, we can also observe that there is a large performance gap between the Any-2 and the All-3 strategy. The reason for this lies in the heterogeneous setup we use. The two organizations running PostgreSQL are able to process the workload significantly faster than the single organization running MySQL. Thus, under the Any-2 strategy, the two organizations using PostgreSQL are able to lead the progress, without having to wait for the significantly slower third organization. Under the All-3 strategy, the progress is throttled by the slowest organizations running MySQL. 

The difference in processing speed also becomes visible, if we inspect the throughput of the stand-alone single-instance database systems in Figure~\ref{figs:throughput2} under the same workload. This time, we fire the transactions using OLTP-bench~\cite{oltp_bench}. Note that both system are configured with a buffer size of $2$GB to keep the working set in main memory.
As we can see, PostgreSQL significantly outperforms MySQL under this workload independent of the number of clients. 

There is one more observation we can make: By comparing Figure~\ref{figs:throughput1} and Figure~\ref{figs:throughput2} side-by-side, we can see that ChainifyDB introduces only negligible overhead over the raw database systems. In fact, for $3$, $6$, and $12$~clients, ChainifyDB under the Any-2 policy actually produces a slightly higher throughput than raw PostgreSQL. The reasons for this lies in our optimized parallel transaction execution, which exploits the batch-wise inflow of transactions, and executes the transaction at the lowest possible isolation level. 

For completeness, we also show in Table~~\ref{table:throughput_distribution} the throughput for Smallbank, where the accounts are picked following a uniform distribution. As we can see, the throughput under a uniform distribution is even higher with up to $6144$~transactions per second than under the skewed Zipf distribution, as it allows for a higher degree of parallelism during execution due to less conflicts between transactions. 

\begin{table}[!htb]
	\setlength{\tabcolsep}{3pt}
	\begin{center}
		\begin{tabular}{p{4.5cm} || p{2.7cm} | p{2.5cm} | p{2.5cm} | p{2.5cm} }
			\hline
			 \textbf{Distribution}& \textbf{3 Clients} & \textbf{6 Clients} & \textbf{12 Clients} & \textbf{24 Clients}\\
			\hline
			Zipf & $2757$ TPS & $3676$ TPS & $4709$ TPS & $4812$ TPS\\\hline
			Uniform & $2279$ TPS & $3840$ TPS & $5774$ TPS & $6144$ TPS \\
			\hline
		\end{tabular}
	\end{center}
	\caption{Average throughput of successful transactions  for ChainifyDB (Any-2) under Smallbank following a Zipf distribution and a uniform distribution.}
	\label{table:throughput_distribution}
\end{table}

\subsection{Robustness and Recovery}
\label{ssec:robustness}

\begin{figure}[!htb]
\begin{center}
\subfigure[Heterogeneous Setup (2x PostgreSQL, 1x MySQL).] {
\includegraphics[width=.7\linewidth, page=6, trim={3cm 2cm 4cm 5cm}, clip]{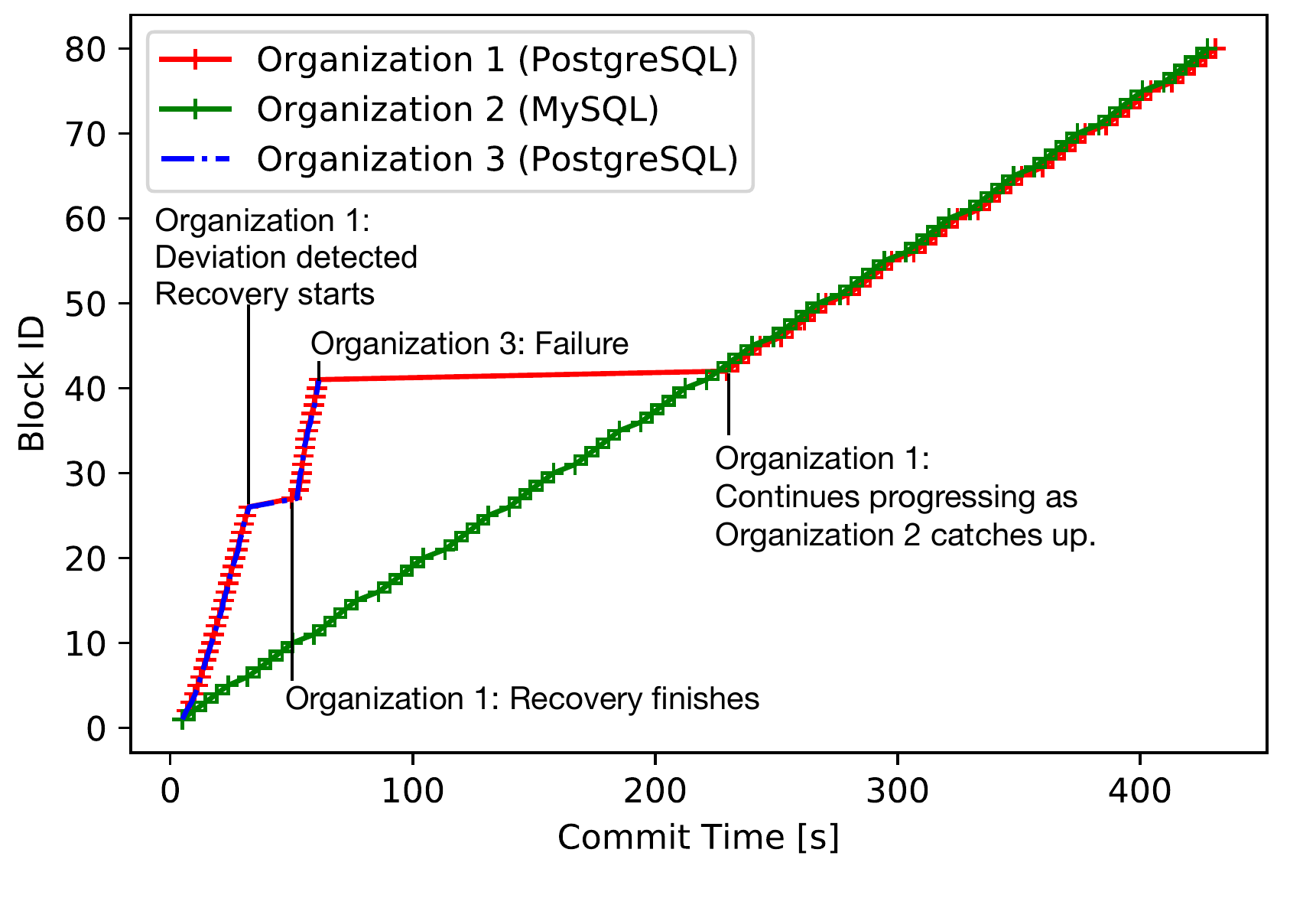}
\label{figs:robustness1}
}
\subfigure[Homogeneous Setup (3x PostgreSQL).] {
\includegraphics[width=.7\linewidth, page=5, trim={3cm 2cm 4cm 5cm}, clip]{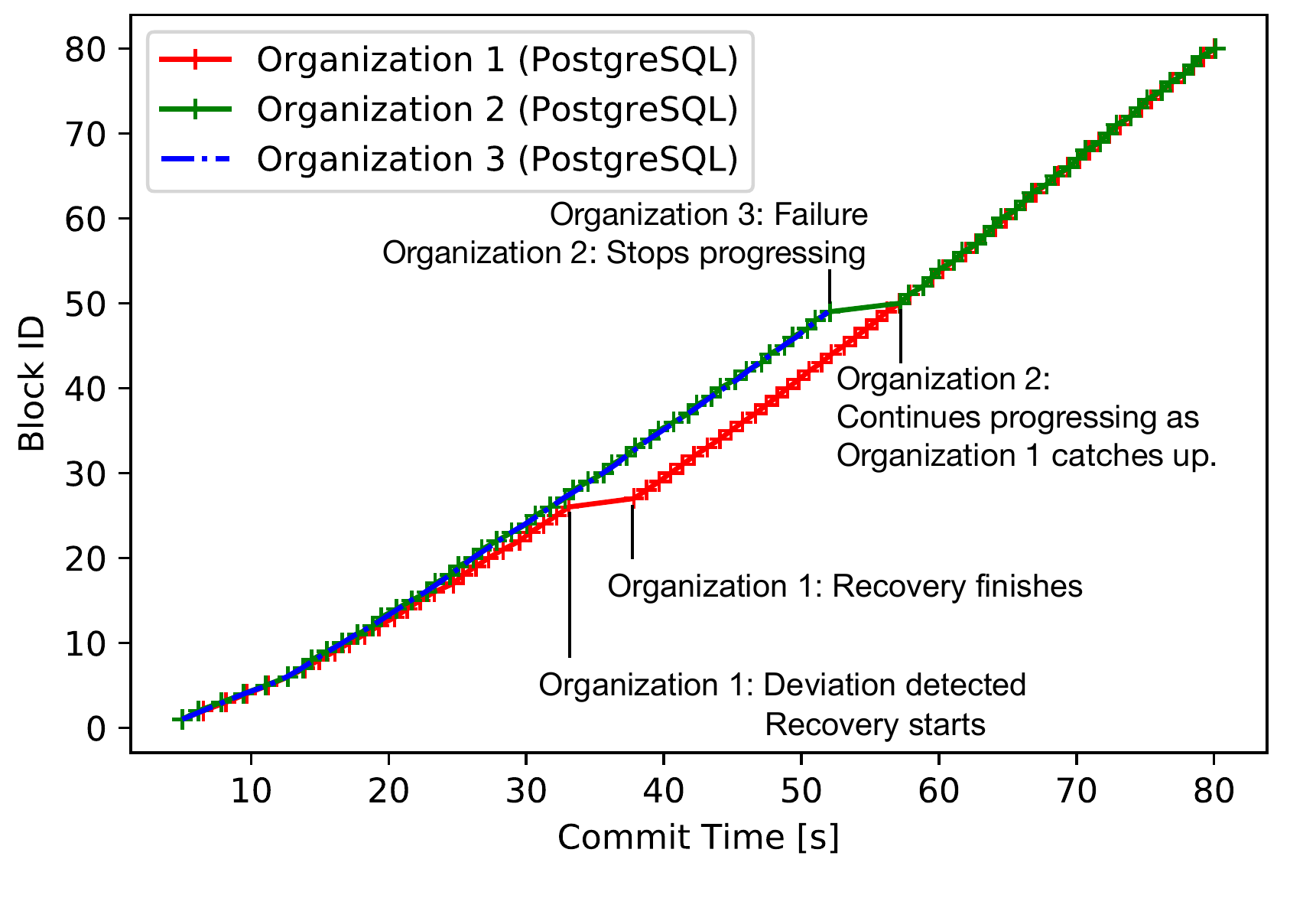}
\label{figs:robustness2}
}
\end{center}
\caption{Robustness and Recovery of ChainifyDB under the Any-2 consensus policy.}
\label{figs:robustness}
\end{figure}

Apart from the transaction processing performance, the robustness and recovery capabilities are crucial properties of ChainifyDB as well. To put these capabilities to the test, in the following experiment, we will disturb our ChainifyDB network in two different ways: First, we forcefully corrupt the database of one organization and see whether ChainifyDB is able to detect and recover from it. Afterwards, we bring down one organization entirely and see whether the network is able to continue progressing. Of course, we are also interested in the performance of the recovery processes.
  
Precisely, we have the following setup for this experiment: In the first phase, we sustain our ChainifyDB network with transactions of the Smallbank workload. These do not cause the organizations to deviate in any way. Consequently, this phase essentially resembles the standard processing situation of ChainifyDB. 
Then, after a certain amount of time, we manually inject an update to the table of organization~$O_1$ and see how fast~$O_1$ is able to recover from the deviation. Note that we do not perform this update through a ChainifyDB transaction, but externally by directly modifying the table in the database. 
Finally, we simulate a complete failure of one organization by removing it from the network. The remaining two organizations then have to reach consensus to be able to progress under the Any-2 policy.  

In Figure~\ref{figs:robustness1}, we visualize the progress of all organizations for our typical heterogeneous setup. Additionally, in Figure~\ref{figs:robustness2}, we test a homogeneous setup, where all three organizations run PostgreSQL. On the $x$-axis, we plot the time of commit for each block. On the $y$-axis, we plot the corresponding block IDs. Every five committed blocks, each organizations creates a local checkpoint. 

Let us start with our typical heterogeneous setup in Figure~\ref{figs:robustness1}. First of all, we can observe that the organizations~$O_1$ and $O_3$, which run PostgreSQL, progress much faster than organization~$O_2$ running MySQL. Shortly after the update has been applied to~$O_1$, it detects the deviation in the consensus round and starts recovery from the most recent checkpoint. Interestingly, this also stops the progression of organization~$O_3$, as $O_3$ is not able to reach consensus anymore according to the Any-2 policy: $O_1$ is busy with recovery and $O_2$ is too far behind. 
As soon as $O_1$ recovers, which takes around~$17$~seconds, $O_3$ also restarts progressing, as consensus can be reached again. 
Both $O_1$ and $O_3$ progress until we let $O_3$ fail. Now, $O_1$ can not progress anymore, as $O_3$ is not reachable and $O_2$ still too far behind due its slow database system running underneath. Thus, $O_1$ halts until $O_2$ has caught up. As soon as this is the case, both $O_1$ and $O_2$ continue progressing at the speed of the slower organization, namely $O_2$. 

In Figure~\ref{figs:robustness2}, we retry this experiment on a homogeneous setup, where all organization run PostgreSQL. Thus, this time there is no drastically slower organization throttling the network. 
Again, at a certain point in time, we externally corrupt the database of organization~$O_1$ by performing an update and $O_1$ starts to recover from the most recent checkpoint. In contrast to the previous experiment, this time the recovery of~$O_1$ does not negatively influence any other organization: $O_2$ and $O_3$ can still reach consensus under the Any-2 policy and continue progressing, as none of the two is behind the other one. Recovery itself takes only around $4$~seconds this time and in this case, another organization is ready to perform a consensus round right after recovery.  
When organization~$O_3$ fails, $O_2$ has to halt processing for a short amount of time, as organization~$O_1$ has to catch up.   

In summary, these experiments show (a)~that we can detect state deviation and recover from it, (b)~that the network can progress in the presence of organization failures, (c)~that all organizations respect the consensus policy at all times, and (d)~that recover neither penalizes the individual organizations nor the entire network significantly.

\subsection{Cost Breakdown}
\label{ssec:breakdown}

In Section~\ref{ssec:throughput}, we have seen the end-to-end performance of ChainifyDB. In the following, we want to analyze how much individual components contribute to this performance.

Precisely, we want to investigate: (a)~The cost of all cryptographic computation, such as signing and validation,  that is happening at several stages in the pipeline (see Section~\ref{sec:arch} for details). (b)~The impact of parallel transaction execution on the underlying database system (see Section~\ref{ssec:parallelism} for details).

\begin{figure}[htb!]
\begin{center}
  \includegraphics[page=4, width=.7\linewidth, trim={1cm 3.5cm 4cm 5cm}, clip]{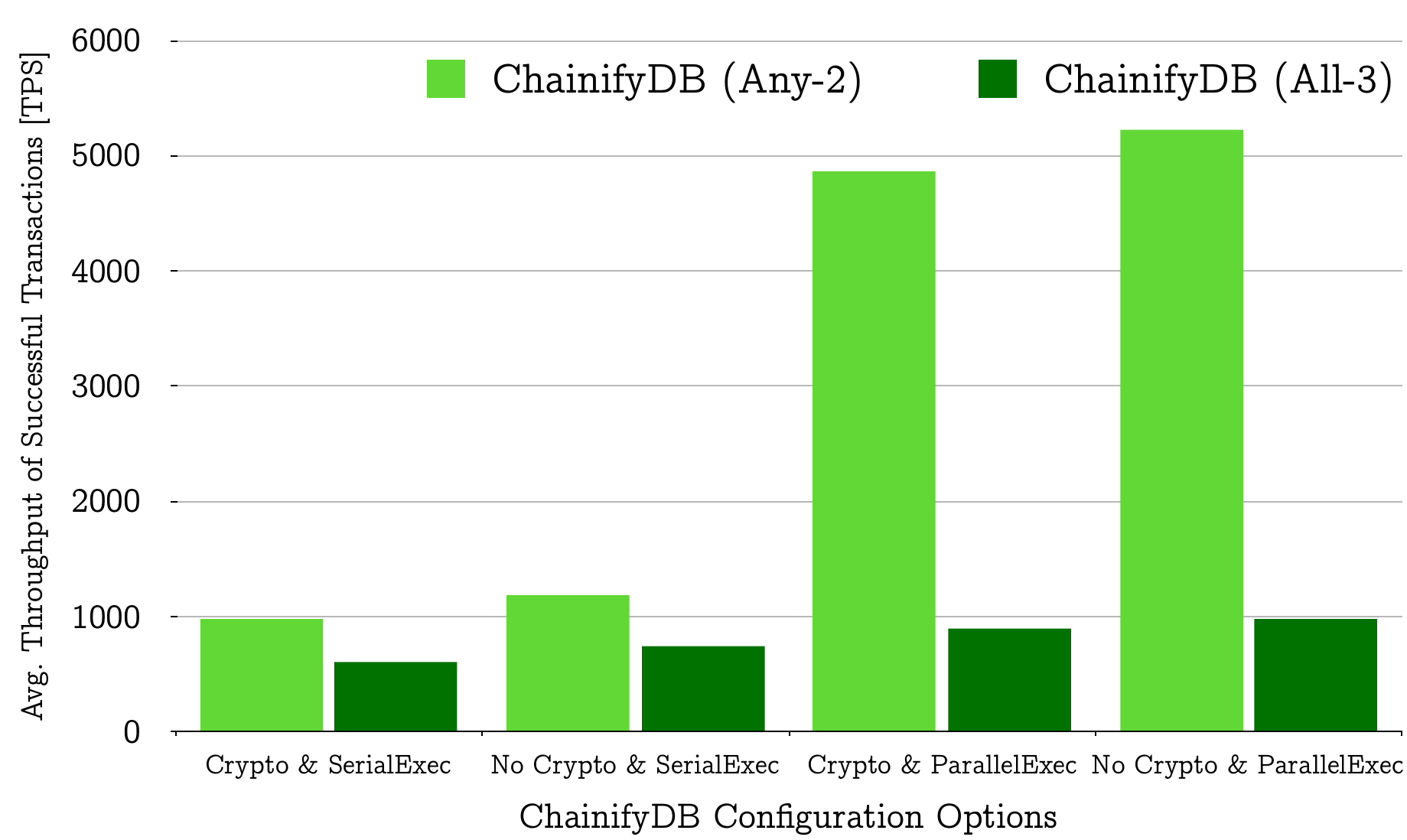}
\end{center}
  \caption{Cost breakdown of ChainifyDB.}
  \label{fig:breakdown}
\end{figure}

Figure~\ref{fig:breakdown} shows the results. 
We can observe that the overhead caused by cryptographic computation is surprisingly small. Under the Any-2 policy, turning on all cryptographic components decreases the throughput only by $7\%$ for parallel execution. Under the All-3 policy, the decrease is only~$8.5\%$.  
While our cryptographic components have little negative effects, our parallel transaction execution obviously has a very positive one. With activated cryptography, parallel transaction execution improves the throughput by up to~$5$x.

\subsection{Varying Blocksize}
\label{ssec:blocksize}
\vspace*{-0.1cm}

Finally, let us inspect the effect of the blocksize, which is an important configuration parameter in any blockchain system. We vary the blocksize from $256$ transactions per block in logarithmic steps up to $4096$ transactions per block and report the average throughput of successful transactions.  

\begin{figure}[htb!]
\begin{center}
  \includegraphics[page=3, width=.7\linewidth, trim={1cm 3cm 1.5cm 5cm}, clip]{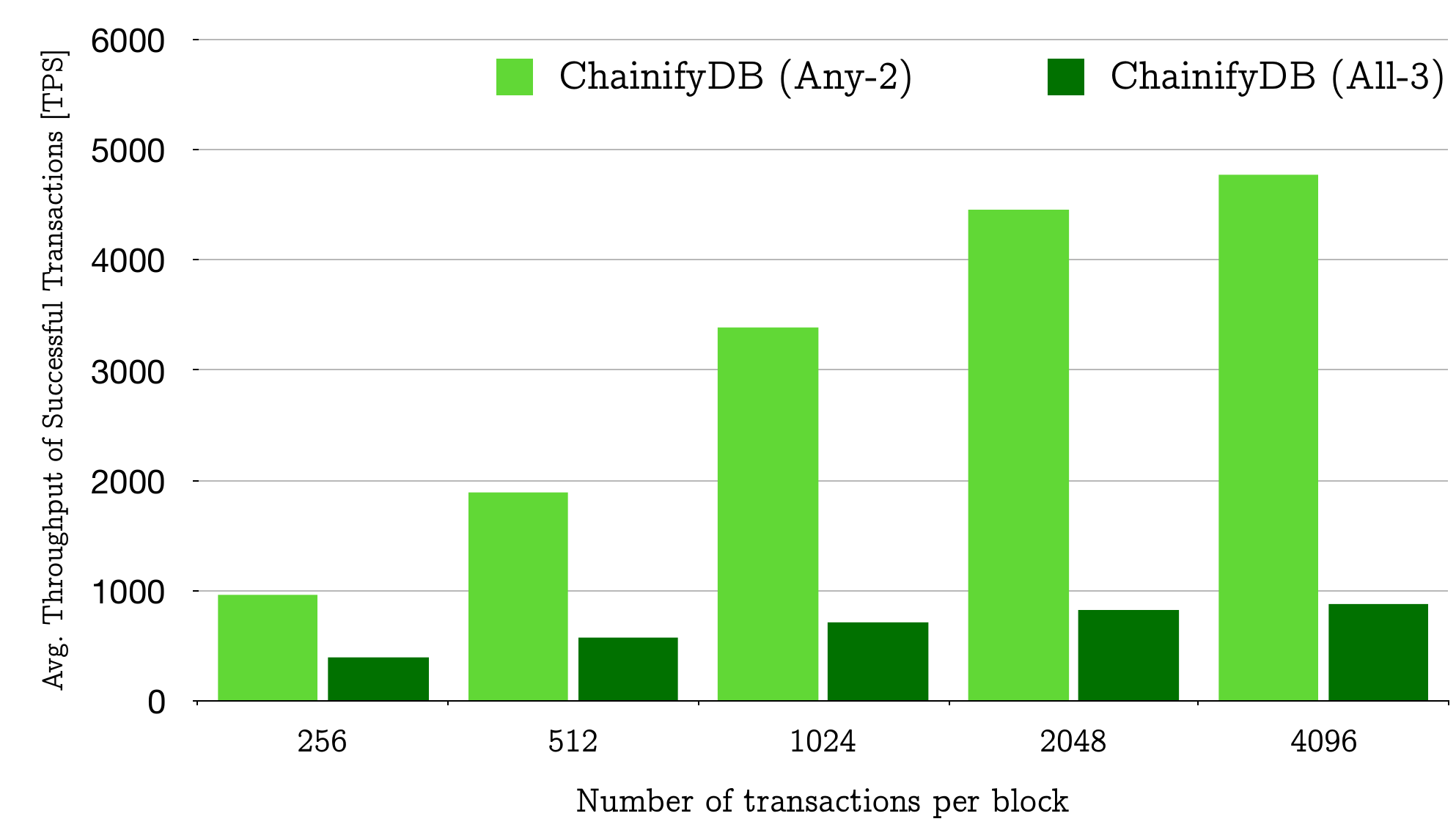}
\end{center}
  \caption{The effect of varying the blocksize.}
  \label{fig:blocksize}
\end{figure}

\noindent Figure~\ref{fig:blocksize} shows the results. We can see that both under the Any-2 and All-3 policy, the throughput increases with the blocksize. This increase is mainly caused by our parallel transaction execution mechanism, which analyzes the whole block of transactions and schedules them for parallel conflict-free execution. 

\section{Conclusion}
In this work, we introduced a highly flexible processing model for permissioned blockchain systems called the Whatever-LedgerConsensus model. WLC avoids making assumptions on the deterministic behavior of individual organizations by reaching consensus on the effects instead of the actions. We clearly formalized WLC and discussed in detail the recovery options in that landscape. To showcase the strengths of WLC, we proposed ChainifyDB, an implementation of a blockchain layer, which is able to chainify arbitrary data management systems and connect them in a network. In an extensive experimental evaluation, we showed that ChainifyDB does not only offer a $6$x~higher throughput than comparable baselines, but also introduces a robust recovery mechanism, which grant organizations the chance to rehabilitate.

\bibliographystyle{abbrv}
\bibliography{chainifydb} 

\newpage
\appendix

\section{Algorithms}

\begin{algorithm}
\lstset{style=MYC}
\begin{lstlisting}[escapeinside={(*}{*)}]
"""
 Function to run consensus for a given block in ChainifyDB.
 
 Parameters:
 	@block_id: The id of the block for which the consensus should run
	@local_hash: The locally computed hash_digest of this block
	@orgs: List of organizations participating in the chainifydb's network
	@policy: Consensus policy specifying the number of organizations that must agree during consensus round
"""
func consensus(block_id, local_hash, orgs[], policy) {
	
	// A hash mapping the hash_digest to the number of
	// organisations who computed this hash_digest
	digest_map := map[key: digest, value: count]
	
	// insert local hash_digest into the map
	digest_map[local_hash_digest] = 1
	
	// get every organisation's hash_digest for this block
	foreach org in organisations {

		// get hash_digest and signature
		hash_digest, sign = org.get_hash_digest(block_id)
		
		// verify if the digest is properly signed
		if sign.Verify(hash_digest, signature, org.PubKey) {
			digest_map[hash_digest] += 1
		}
	}
	
	// check if the consensus is reached
	for hash, count in digest_map {
		if count >= policy.K && hash == local_hash {
			// consensus has been reached
			// and local node is consenting
			return hash, true
		} else {
			// consensus reached
			// but this node is non-consenting
			return hash, false  // start recovery
		}
	}
	
	// no consensus has been reached
	return nil, false
}
\end{lstlisting}
\caption{ChainifyDB consensus mechanism executed by each organization individually.}
\label{alg:consensus}
\end{algorithm}

\begin{algorithm}[ht!]
\lstset{style=MYC}
\begin{lstlisting}[escapeinside={(*}{*)}]
"""
 Function to perform recovery of an organization.
 
 Parameters:
 	@non_consenting_block_id: 
	   The id of the block that caused non-consenting
	@blocks: 
	   The history of blocks up to the non-consenting one 
	@table: The table that is non-consenting
	@local_snapshots: 
	   The locally committed snapshots, ordered from newest to oldest
"""
func recoverNode(non_consenting_block_id, 
                 blocks, 
                 table, 
                 local_snapshots) {
   // Assumption: the last block executed by this 
   // organization was non-consenting.
   
   // try restoring from local snapshots from newest 
   // to oldest 
   while(local_snapshots) {
      // get newest snapshot
      snapshot = local_snapshots.popNewest()
      // copy snapshot to table
      table = snapshot
      // replay all blocks up to (including) the 
      // non-consenting one
      replay(table, blocks, snapshot.block_id + 1, 
                            non_consenting_block_id)
      // if the table is now consenting, we are done
      if(isConsenting(table)) return true
   }

   // non available snapshot lead to a consenting table
   // thus, replay the entire history
   replay(table, blocks, 0, non_consenting_block_id)
   if(isConsenting(table)) return true
  
   // the organization is still non-consenting
   // so exclude it from the network
   return false
}

// helper procedure to replay all blocks specified 
// in a range
func replay(table, blocks, begin_block_id, end_block_id){
   for(b = begin_block_id; b <= end_block_id; b++) {
      table.execute(blocks[b])
   }
}
\end{lstlisting}
\caption{ChainifyDB's recovery algorithm in pseudo-code.}
\label{alg:recovery}
\end{algorithm}

\end{document}